\documentclass[aps,prd,onecolumn,nofootinbib,showkeys]{revtex4-1} 

\usepackage[T1]{fontenc}
\usepackage[utf8]{inputenc}
\usepackage{amsmath,amssymb,amsfonts}
\usepackage{graphicx,epsfig}
\usepackage{color}
\usepackage{hyperref}
\usepackage{array}
\usepackage{subfigure}  
\usepackage{multirow}
\usepackage{float}
\usepackage{booktabs}

\begin{document}

\title{Charged Black Holes in the Kalb-Ramond Background with Lorentz Violation: \\
Null Geodesics and Optical Appearance of a Thin Accretion Disk}

\author{K.TAN}
\email{tk14708229726@163.com}
\author{X.G.Lan}
\email{Corresponding author:xglan@cwnu.edu.cn}
\affiliation{ Institute of Theoretical Physics, China West Normal University, Nanchong 637009, China}

\date{\today}

\begin{abstract}

\textbf{Abstract:} In this paper, we investigate the optical appearance of a charged black hole in the Kalb-Ramond background, incorporating a Lorentz-violating parameter $l=0.01$. By analyzing the null geodesics, we derive the photon sphere, event horizon, effective potential, and critical impact parameters. We then employ a ray-tracing technique to study the trajectories of photons surrounding a thin accretion disk. Three different emission models are considered to explore the observed intensity profiles of direct rings, lensing rings, and photon sphere. By comparing these results with those of the standard Reissner-Nordström black hole ($l=0$) and the Kalb-Ramond black hole with different values of Lorentz-violating parameter  (specifically, $l=0.05$ and $l=0.1$ respectively), we find that the Lorentz symmetry breaking will lead to a decrease in the radii of the photon sphere, the event horizon, and the innermost stable circular orbit. Consequently, this makes the detection of these black holes more challenging.

\end{abstract}

\maketitle
\tableofcontents  

\section{Introduction}
Before the Event Horizon Telescope (EHT) revealed images of supermassive black holes, Karl Schwarzschild provided the first exact solution to Einstein’s field equations in 1916~\cite{Schwarzschild:1916uq}. In 1963, Roy Kerr introduced the precise solution describing a rotating black hole~\cite{Kerr:1963ud}. The release of images depicting the supermassive black holes in M87* and our Milky Way's center, Sgr A*, has confirmed the existence of black holes~\cite{Narayan:2019imo,Zakharov:2023yjl,Safarzadeh:2019imq}. These images show a central dark region, referred to as the black hole shadow, along with a bright photon sphere~\cite{Zeng:2020vsj,He:2024bll,Wei:2021vdx}. Synge first proposed the concept of the black hole shadow in 1966~\cite{Synge:1966okc}. The observed dark disk encircled by a bright ring is attributed to the powerful gravitational lensing near the black hole~\cite{Bardeen:1973tla}. Photons falling within the shadow are absorbed, whereas photons orbiting near the shadow may reach a distant observer multiple times and manifest as part of the bright photon ring. This ring is often deemed the boundary of the black hole horizon or the critical curve~\cite{He:2022yse,He:2021htq}. In the Schwarzschild spacetime, the bound photon orbit is located at $r=3M$, corresponding to an impact parameter $b=3\sqrt{3}M$. Consequently, the black hole shadow is demarcated by the region inside this parameter.

Observations and theoretical research on black holes continue to expand~\cite{Perlick:2021aok,Jin:2020emq,Guo:2020zmf,Chen:2019fsq,Mustafa:2022xod}. When analyzing black hole shadows, the accretion flow plays a decisive role in visibility. A variety of methods focus on how the shadow structure is influenced by the accretion disk's geometry and optical depth~\cite{Zeng:2021mok,Yang:2024nin,He:2024amh}. Many studies utilize a thin accretion disk, as employed by Gralla \emph{et al}.~\cite{Gralla:2019xty}, who classified photon trajectories near the shadow into direct rings, lensing rings, and photon sphere. Though realistic accretion streams are usually not spherically symmetric, idealized spherical models help elucidate the black hole shadows. Various corrections to general relativity in modified gravity theories can alter the shadow’s size and shape, facilitating novel tests of gravity. Additionally, black hole parameters such as charge and spin have a substantial effect on shadows~\cite{Li:2021ypw,Bronzwaer:2021lzo,Gan:2021xdl,EventHorizonTelescope:2021dqv,Gan:2021pwu,Wang:2023vcv,Uniyal:2023inx}.

Black holes in the Kalb-Ramond background present an interesting scenario for exploring possible Lorentz symmetry violations~\cite{Junior:2024ety,Nandi:2023vxt,Jusufi:2017hed,Zhang:2025wnu,Kanzi:2019gtu}. Such a background may modify the spacetime geometry, black hole thermodynamics, and the behavior of matter and radiation around the event horizon~\cite{Ditta:2024lnb,Ma:2024ets}. It may also alter the silhouette of the black hole shadow~\cite{Zahid:2024hyy,Jumaniyozov:2025wcs,al-Badawi:2024pdx,Liu:2024oas,Raza:2024zkp,Ortiqboev:2024mtk}. Because the Kalb-Ramond field emerges naturally in string theory, its study can deepen our understanding of fundamental physics~\cite{Kumar:2020hgm,Junior:2024vdk,Atamurotov:2022wsr,Mavromatos:2018drr}. Motivated by these considerations, we focus on examining the shadows and observational characteristics of a charged black hole in the Kalb-Ramond background, with the Lorentz violation parameter set to $l=0.01$. We study the observed intensity of the shadow under various charge values and thin accretion disk emission models.

This paper is structured as follows. Section~\ref{sec:Shadow} derives the null geodesics of a charged black hole in the Kalb-Ramond background and explores the corresponding photon sphere radii. Section~\ref{sec:QNMs} applies a ray-tracing technique to evaluate the observational features of a thin accretion disk for different emission models. Section~\ref{sec:Comparison} compares our results with the standard Reissner-Nordström black hole to highlight the effects of Lorentz violation. Section~\ref{sec:Conclusion} summarizes our main findings.

\section{The null geodesics of a charged black hole in the Kalb-Ramond background}\label{sec:Shadow}
The study of a charged black hole in the Kalb-Ramond field within a spherically symmetric background has a rich theoretical context, in general relativity. The Kalb-Ramond field, originally introduced in string theory as a two-form gauge field, can have significant implications for the properties of black holes. We start from the Kalb-Ramond field action~\cite{Lessa:2019bgi,Altschul:2009ae}
\begin{equation}\label{eq1}
\begin{aligned}
S & =\frac{1}{2} \int d^4 x \sqrt{-g}\left[R-2 \Lambda-\frac{1}{6} H^{\mu \nu \rho} H_{\mu \nu \rho}-V\left(B^{\mu \nu} B_{\mu \nu} \pm b^2\right)+\xi_1 B^{\rho \mu} B_\mu^\nu R_{\rho \nu}+\xi_2 B^{\mu \nu} B_{\mu \nu} R\right] \\
& +\int d^4 x \sqrt{-g} \mathcal{L}_{\mathrm{M}},
\end{aligned}
\end{equation}
where
\begin{equation}
B^{\mu \nu} B_{\mu \nu}=\mp b^2,
\end{equation}
\begin{equation}
\mathcal{L}_{\mathrm{M}}=-\frac{1}{2} F^{\mu \nu} F_{\mu \nu}-\eta B^{\alpha \beta} B^{\gamma \rho} F_{\alpha \beta} F_{\gamma \rho},
\end{equation}
\begin{equation}
F_{\mu \nu}=\partial_\mu A_\nu-\partial_\nu A_\mu,
\end{equation}
\begin{equation}
\tilde{H}^{\mu \nu \rho} \tilde{H}_{\mu \nu \rho}=H^{\mu \nu \rho} H_{\mu \nu \rho}+2 H^{\mu \nu \rho} A_{[\mu} F_{\nu \rho]}+A^{[\mu} F^{\nu \rho]} A_{[\mu} F_{\nu \rho]}.
\end{equation}
Here, $R$ refers to the Ricci scalar, $\eta$ is the coupling constant, $\xi_{1,2}$ denotes the non-minimal coupling constant between gravity and the Kalb-Ramond field, and $\mathcal{L}_{\mathrm{M}}$ represents the Lagrangian of the electromagnetic field. To maintain invariance under Lorentz transformations, we assume that $V\left(B^{\mu \nu} B_{\mu \nu} \pm b^2\right)$ depends on $B^{\mu \nu} B_{\mu \nu}$.

The modified Einstein equation is obtained by varying the action with respect to the metric $g^{\mu \nu}$, and is given by the following form:
\begin{equation}
R_{\mu \nu}-\frac{1}{2} g_{\mu \nu} R+\Lambda g_{\mu \nu}=T_{\mu \nu}^{\mathrm{M}}+T_{\mu \nu}^{\mathrm{KR}}.
\end{equation}
Here, $T_{\mu \nu}^{\mathrm{M}}$ is the energy-momentum tensor of the electromagnetic field, and $T_{\mu \nu}^{\mathrm{KR}}$ is the effective energy-momentum tensor of the Kalb-Ramond field.

In the vacuum, the modified Einstein equation can be expressed as
\begin{equation}
\begin{aligned}
R_{\mu \nu} & =T_{\mu \nu}^{\mathrm{M}}-\frac{1}{2} g_{\mu \nu} T^{\mathrm{M}}+\Lambda g_{\mu \nu}+V^{\prime}\left(2 b_{\mu \alpha} b_\nu{ }^\alpha+b^2 g_{\mu \nu}\right) \\
& +\xi_2\left[g_{\mu \nu} b^{\alpha \gamma} b^\beta{ }_\gamma R_{\alpha \beta}-b^\alpha{ }_\mu b^\beta{ }_\nu R_{\alpha \beta}-b^{\alpha \beta} b_{\mu \beta} R_{\nu \alpha}-b^{\alpha \beta} b_{\nu \beta} R_{\mu \alpha}\right. \\
& \left.+\frac{1}{2} \nabla_\alpha \nabla_\mu\left(b^{\alpha \beta} b_{\nu \beta}\right)+\frac{1}{2} \nabla_\alpha \nabla_\nu\left(b^{\alpha \beta} b_{\mu \beta}\right)-\frac{1}{2} \nabla^\alpha \nabla_\alpha\left(b_\mu{ }^\gamma b_{\nu \gamma}\right)\right].
\end{aligned}
\end{equation}
When the cosmological constant $\Lambda$ is set to zero, the metric describing a charged black hole in a spherically symmetric spacetime within the framework of the Kalb-Ramond field can be expressed as follows~\cite{Duan:2023gng}:
\begin{equation}\label{eq99}
d S^2=-F(r) \, d t^2+\frac{1}{F(r)} \, d r^2+r^2\left(d \theta^2+\sin ^2 \theta \, d \varphi^2\right),
\end{equation}
where
\begin{equation}\label{eq2}
F(r)=\frac{1}{1-l}-\frac{2 M}{r}+\frac{Q^2}{(1-l)^2 r^2}.
\end{equation}
Here, $M$ and $Q$ denote the mass parameter and the charge parameter, respectively. The Kalb-Ramond field originates from the rank-two antisymmetric tensor field in string theory, and the introduction of the Lorentz-violating parameter $l$ in Eq .~(\ref{eq2}) is precisely derived from the Kalb-Ramond field. In the context of Lorentz symmetry violation, this parameter is used to modify the geometric properties and topological characteristics of spacetime. The strength of the Lorentz-violating effects is represented by the parameter $l$ in Eq .~(\ref{eq2}). Such modifications are consistent with the framework of effective field theory, as similar terms can arise from higher-dimensional operators or non-minimal couplings~\cite{Kumar:2020hgm,Junior:2024ety,Atamurotov:2022wsr}.

In string theory, it is known that the Kalb-Ramond field couples to the spacetime metric of black holes, potentially leading to Lorentz symmetry-violating corrections. These corrections influence the black hole solutions, and as a result, the Kalb-Ramond field may affect key features of spacetime, such as the event horizon, photon sphere, and other important characteristics. To ensure that the Lorentz-violating effects remain perturbative and do not significantly alter the geometric structure of the black hole, the parameter in Eq .~(\ref{eq2}) is set to a small value ($l=0.01$)~\cite{Kumar:2020hgm,Junior:2024ety,Atamurotov:2022wsr}. 

When $l=0$, this metric reduces to the Reissner-Nordström black hole. Further setting $Q=0$ recovers the Schwarzschild black hole. Throughout this work, we focus on the case $l=0.01$, leading to
\begin{equation}
F(r)=\frac{1}{1-0.01}-\frac{2 M}{r}+\frac{Q^2}{(1-0.01)^2\,r^2}.
\end{equation}

The horizon radius of this charged black hole follows from
\begin{equation}\label{eq4}
F(r)=0.
\end{equation}
The largest positive root of Eq.~(\ref{eq4}) gives the event horizon. To ensure the black hole solution exists, $Q$ must satisfy $\frac{Q^2}{M^2} \leq (1-l)^3$. Therefore, we restrict $0 \leq Q \leq 0.6$.

Subsequently, we investigate the photon motion via the Euler-Lagrange equation
\begin{equation}\label{eq5}
\frac{d}{d \lambda}\left(\frac{\partial \mathcal{L}}{\partial \dot{x}^\mu}\right)=\frac{\partial \mathcal{L}}{\partial x^\mu},
\end{equation}
where $\lambda$ is the affine parameter and $\dot{x}^\mu$ denotes the photon’s four-velocity. The Lagrangian $\mathcal{L}$ is
\begin{equation}
\mathcal{L}=\frac{1}{2} g_{\mu \nu}\frac{d x^\mu}{d \lambda}\frac{d x^\nu}{d \lambda}
=\frac{1}{2}\left[-F(r)\,\dot{t}+\frac{1}{F(r)}\,\dot{r}+r^2(\dot{\theta}+\sin^2 \theta \,\dot{\varphi}^2)\right].
\end{equation}
Because the spacetime is spherically symmetric, we restrict photon motion to the equatorial plane $\theta=0$, where $\dot{\theta}=0$. The time-translation and $\alpha$-rotational symmetries yield two conserved quantities
\begin{equation}\label{eq14}
E=F(r)\,\dot{t}, 
\quad 
L=r^2 \,\dot{\varphi}.
\end{equation}
The modification introduced by the Kalb-Ramond background mainly affects the conserved quantities of the system in terms of the parameter $l$ . As can be seen from Eq.~(\ref{eq14}), the parameter $l$ has an impact on the conserved quantity $E$, but the parameter $l$ has no effect on the conserved quantity $l$.

Generally speaking, the Hamiltonian form of the equation of motion is $\dot{x}^\mu=\frac{\partial H}{\partial p_\mu}, \dot{p}_\mu=-\frac{\partial H}{\partial x^\mu}$. By combining the corresponding conserved quantities and the specific spacetime background, we can obtain an equation of motion similar to that in the paper. For an axisymmetric spacetime, it is relatively more complex, and the Carter constant needs to be introduced to obtain the analytical form of the equation of motion. In this paper, it is a spherically symmetric spacetime, which is relatively simple. By combining the conserved quantities $E$ and $L$, as well as the null geodesic equation, the equation of motion of photons on the equatorial plane can then be described as
\begin{equation}\label{eq8}
\begin{aligned}
\frac{d t}{d \lambda} & =\frac{1}{b_c}\left[\frac{1}{1-0.01}-\frac{2 M}{r}+\frac{Q^2}{(1-0.01)^2\,r^2}\right], \\
\frac{d \varphi}{d \lambda} & = \pm \frac{1}{r^2}, \\
\frac{d r}{d \lambda} & =\sqrt{\frac{1}{b_c^2}-\frac{1}{r^2}\left(\frac{1}{1-0.01}-\frac{2 M}{r}
+\frac{Q^2}{(1-0.01)^2\,r^2}\right)},
\end{aligned}
\end{equation}
where ``$\pm$'' indicates the motion direction. We define the impact parameter as $b_c=\tfrac{L}{E}=\tfrac{r^2\dot{\varphi}}{F(r)\dot{t}}$.

From Eq.~(\ref{eq8}), we have
\begin{equation}\label{eq9}
\dot{r}^2=\frac{1}{b_c^2}-V_{e\!f\!f},
\end{equation}
with
\begin{equation}
V_{e\!f\!f}=\frac{1}{r^2}\left(\frac{1}{1-l}-\frac{2 M}{r}+\frac{Q^2}{(1-l)^2\,r^2}\right).
\end{equation}

\begin{figure*}[htbp]
	\centering
	\subfigure[The effective potential $V_{e\!f\!f}$ varies with the charge $Q$.]{
		\begin{minipage}[t]{0.33\linewidth}
			\includegraphics[width=2.4in]{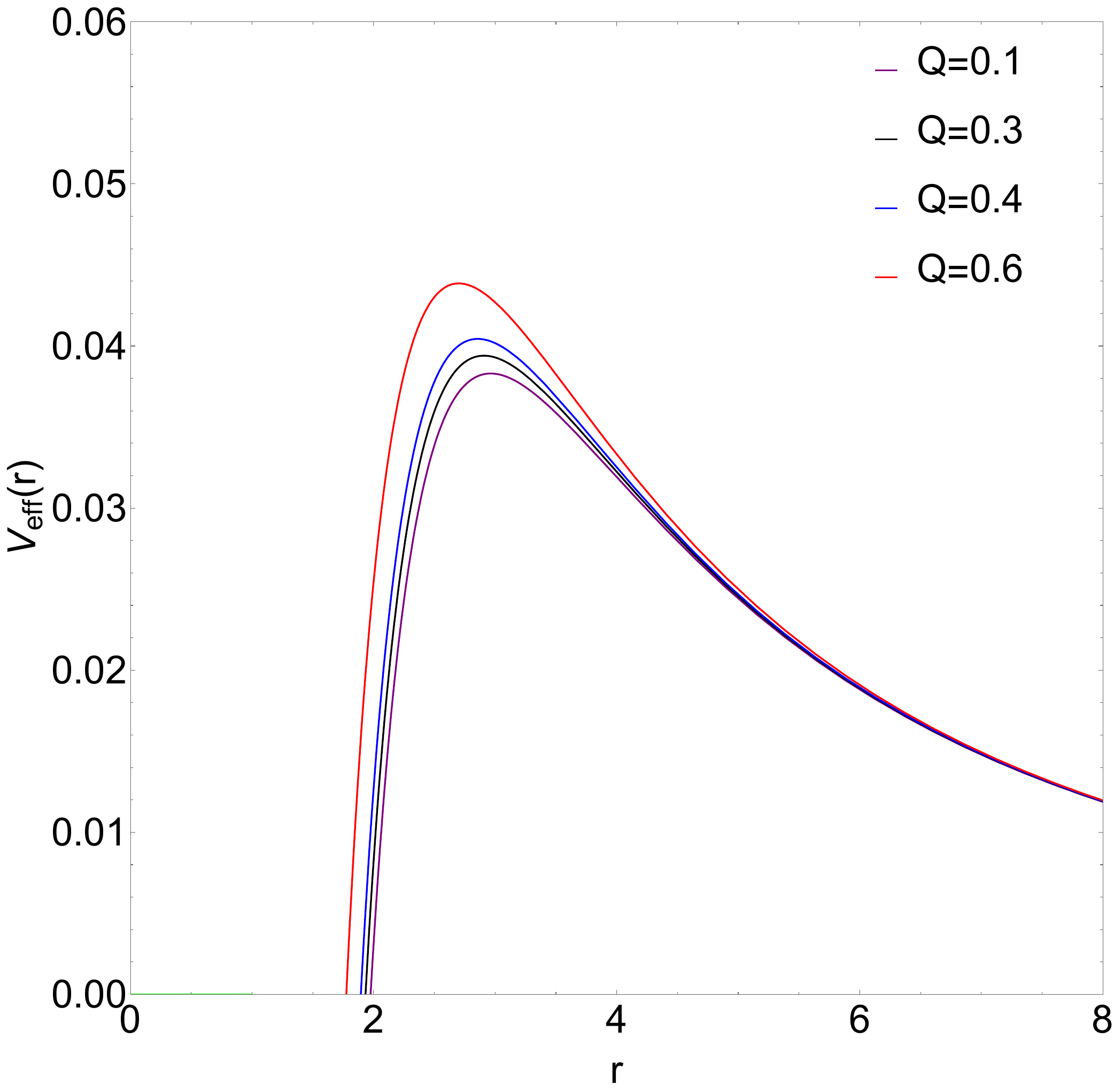}\\
		\end{minipage}
	}
	\subfigure[The effective potential $V_{e\!f\!f}$ varies with the Lorentz-violating parameter $l$.]{
		\begin{minipage}[t]{0.33\linewidth}
			\includegraphics[width=2.4in]{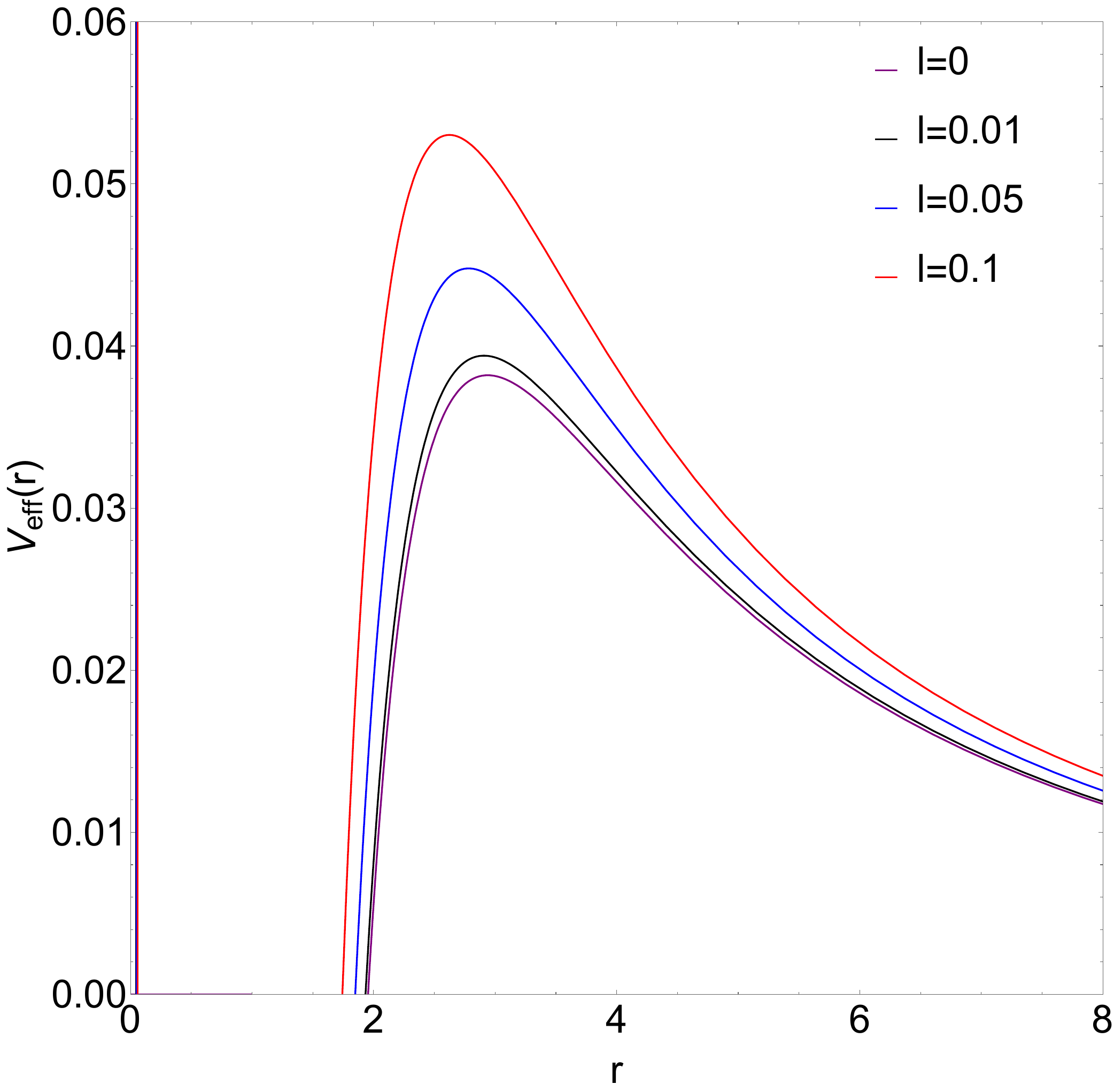}\\
		\end{minipage}
	}
	\caption{(color online) With the mass $M$ set to 1, the contours of the effective potential $V_{e\!f\!f}$ and the impact parameter $b_c$ are presented.}
	\label{fig.veff}
\end{figure*}

The photon sphere radius $r_p$ and the critical impact parameter $b_h$ (shadow radius as seen by a distant observer) satisfy
\begin{equation}\label{eq11}
V_{e\!f\!f}(r_p)=\frac{1}{b_h^2}, 
\quad 
\frac{d}{dr}V_{e\!f\!f}(r_p)=0.
\end{equation}
As shown in Fig.\ref{fig.veff}, the effective potential $V_{e\!f\!f}$ gradually increases from the event horizon radius $r_h$, reaches its peak at the photon sphere radius $r_p$, and then begins to decrease. Both plots in Fig.\ref{fig.veff} illustrate that as the charge $Q$ and Lorentz-violating parameter $l$ increase, the peak value of the effective potential $V_{e\!f\!f}$ increases, and meanwhile, the radius of the photon sphere shrank slightly. We find that the Lorentz-violating parameter $l$ only alters the effective potential while it does not transform the conditions for the stability of the photon sphere.

By effectuating a Taylor expansion of the metric function presented in Eq.~(\ref{eq2}), we derive
\begin{equation}\label{eq19}
F(r)=\left(1+\frac{Q^2}{r^2}-\frac{2 M}{r}\right)+\left(1+\frac{2 Q^2}{r^2}\right) l+\left(1+\frac{3 Q^2}{r^2}\right) l^2.
\end{equation}
Through the combination of Eq.~(\ref{eq11}) and ~(\ref{eq19}), we can obtain the analytical expressions for $r_p$ and $b_h$
\begin{equation}
r_p=\frac{3 M+\sqrt{9 M^2-4\mathcal{A}\left(2 Q^2+4 Q^2 l+6 Q^2 l^2\right)}}{2\mathcal{A}},
\end{equation}
\begin{equation}
b_h=\frac{\sqrt{-27 M^4+36 M^2 Q^2 \mathcal{B}-8 Q^4 \mathcal{B}^2-9 M^3 \sqrt{9 M^2-8 O^2 \mathcal{B}}+8 M Q^2 \mathcal{B} \sqrt{9 M^2-8 Q^2 \mathcal{B}}}}{\sqrt{2} \sqrt{\mathcal{A}\left(-M^2+Q^2\mathcal{B}\right)}},
\end{equation}
where
\begin{equation}
\mathcal{A}=1+l+l^2,  \mathcal{B}=1+3 l+6 l^2+5 l^3+3 l^4.
\end{equation}
Through the photon sphere radius $r_p$ and Eq.~(\ref{eq19}), we can obtain the shadow radius $r_s$ as
\begin{equation}
r_s=\frac{r_p}{\sqrt{F(r_p)}}
=\frac{\left(3 M+\sqrt{9 M^2-8 Q^2 \mathcal{B}}\right) \sqrt{\frac{-9 M^2+4 Q^2 \mathcal-3 M \sqrt{9 M^2-8 Q^2\mathcal{B}}}{\mathcal{A}\left(-3 M^2+2 Q^2 \mathcal{B}-M \sqrt{9 M^2-8 Q^2 \mathcal{B}}\right)}}}{2 \mathcal{A}},
\end{equation}
\begin{figure*}[htbp]
	\centering
	\subfigure[The variation of the photon sphere radius $r_p$ with respect to the Lorentz violation parameter $l$.]{
		\begin{minipage}[t]{0.33\linewidth}
			\includegraphics[width=2.4in]{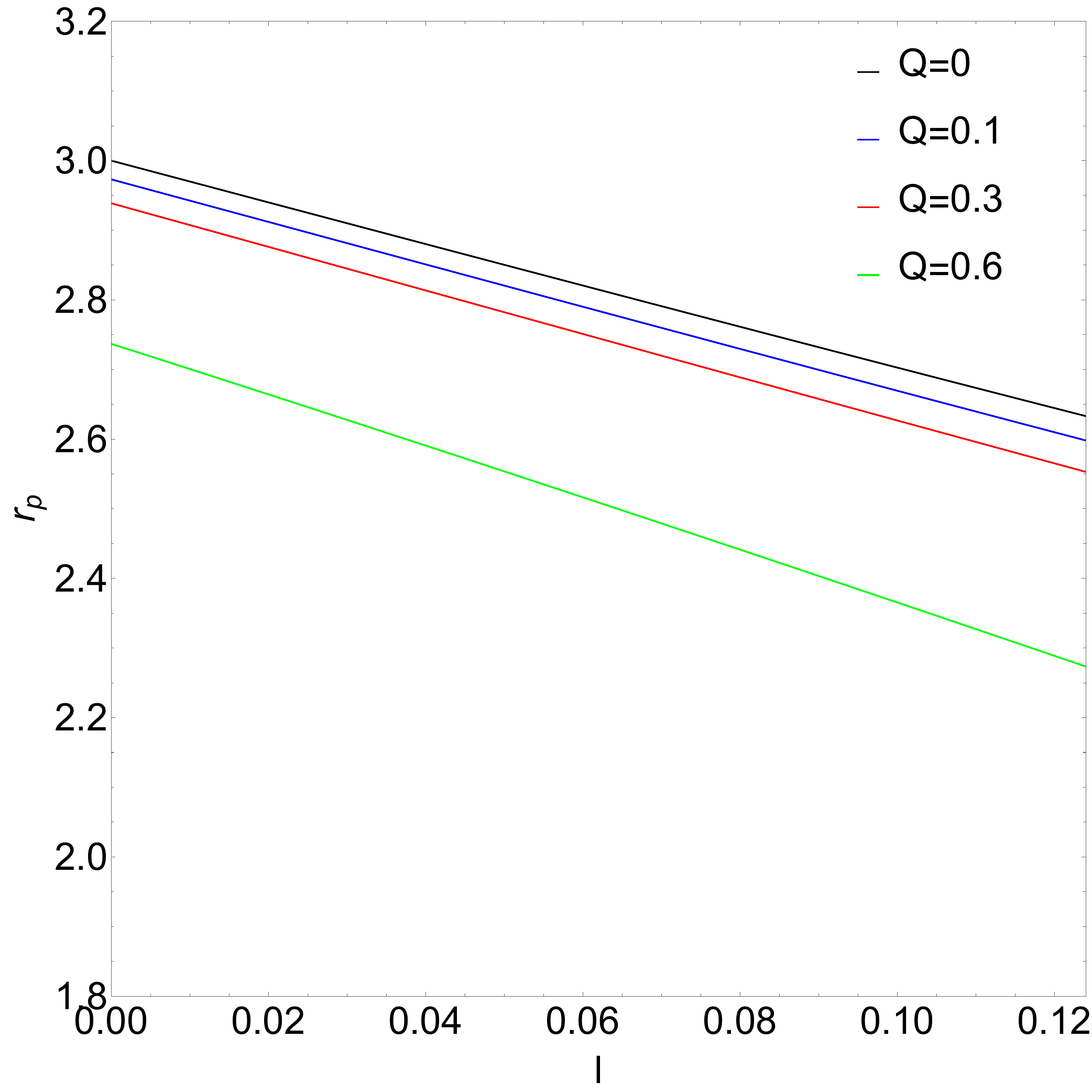}\\
		\end{minipage}
	}
	\subfigure[The variation of the shadow radius $r_s$ with respect to the Lorentz violation parameter $l$.]{
		\begin{minipage}[t]{0.33\linewidth}
			\includegraphics[width=2.4in]{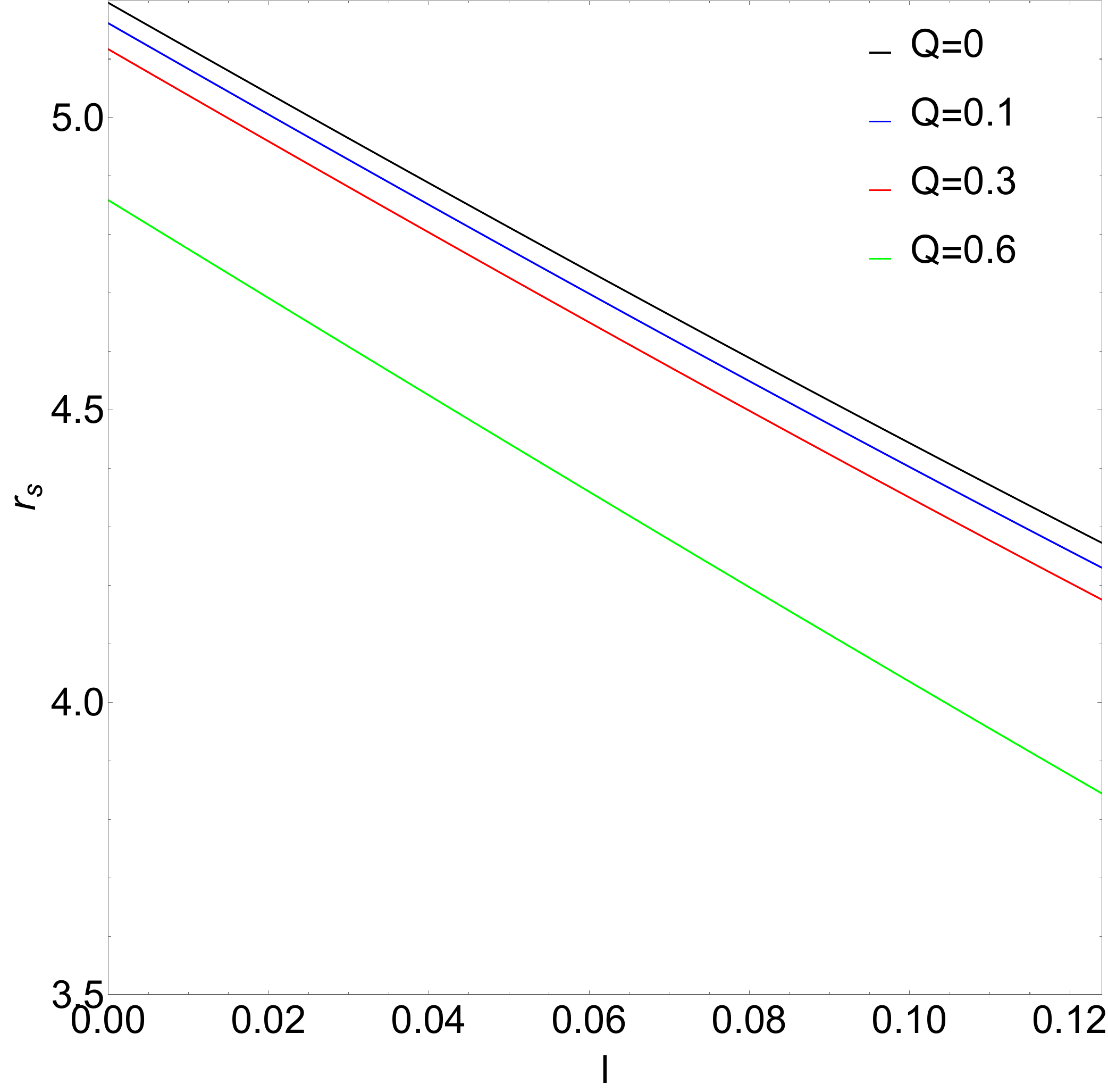}\\
		\end{minipage}
	}
	\caption{(color online) Photon sphere radius $r_p$ and shadow radius $r_s$ as functions of the Lorentz-violation parameter $l$ for a mass $M$ of 1.}
	\label{fig.r}
\end{figure*}

Subsequently, we fix the mass of the black hole at $M = 1$. Under this condition, the photon sphere radius $r_p$, the black hole shadow radius $r_s$, and the critical impact parameter $b_h$ depend on the charge $Q$ and the Lorentz-violating parameter $l$. From Fig.\ref{fig.r}, we observe that as the parameters $l$ and $Q$ increase, they all exhibit a decreasing trend.
As the parameters $l$ and $Q$ increase, they all exhibit a decreasing trend.
\begin{table}[htbp]
\centering
\caption{The variation of $b_h$, $r_p$, and $r_h$ for a standard Kalb-Ramond black hole ($M=1$ and $Q=0.3$) as functions of $l$.}
\label{tan}
\begin{tabular}{cccc}
\toprule
$l$ & $b_h$ & $r_p$ & $r_h$ \\
\midrule
0 & 5.11679 & 2.93875 & 1.95394 \\
0.01 & 5.0378 & 2.90747 & 1.93297 \\
0.05 & 4.72632 & 2.78228 & 1.84876 \\
0.1 & 4.34993 & 2.62677 & 1.74261 \\
\bottomrule
\end{tabular}
\end{table}

\begin{table}[htbp]
\centering
\caption{The critical impact parameter $b_h$, photon sphere radius $r_p$, and event horizon radius $r_h$ for a charged black hole in the Kalb-Ramond background ($l=0.01, M=1$) as functions of $Q$.}
\label{PR}
\begin{tabular}{cccc}
\toprule
$Q$ & $b_h$ & $r_p$ & $r_h$ \\
\midrule
0.1 & 5.1096 & 2.96318 & 1.97489 \\
0.2 & 5.08295 & 2.94254 & 1.95938 \\
0.3 & 5.0378 & 2.90747 & 1.93297 \\
0.4 & 4.9729 & 2.85686 & 1.89470 \\
0.5 & 4.88629 & 2.78891 & 1.84298 \\
0.6 & 4.77487 & 2.70071 & 1.77515 \\
\bottomrule
\end{tabular}
\end{table}

\begin{table}[htbp]
\centering
\caption{The variation of $b_h$, $r_p$, and $r_h$ for a standard Reissner-Nordström black hole ($M=1$) as functions of $Q$.}
\label{PRR}
\begin{tabular}{cccc}
\toprule
$Q$ & $b_h$ & $r_p$ & $r_h$ \\
\midrule
0.1 & 5.18748 & 2.99332 & 1.99499 \\
0.2 & 5.16124 & 2.97309 & 1.97980 \\
0.3 & 5.11679 & 2.93875 & 1.95394 \\
0.4 & 1.91652 & 2.88924 & 5.05298 \\
0.5 & 4.96791 & 4.82288 & 1.86603 \\
0.6 & 4.85869 & 2.73693 & 1.80000 \\
\bottomrule
\end{tabular}
\end{table}

In Tables~\ref{tan} and~\ref{PR}, we calculate a variation patterns of the values of $b_h$, $r_p$, and $r_h$ within the Kalb-Ramond black hole. We can observe that the values of each parameters decrease as both $Q$ and $l$ increase. In Table~\ref{PRR}, we discuss how the values of $b_h$, $r_p$, and $r_h$ change with $Q$ in a Reissner-Nordström black hole. We also find a similar result, that is, each of these parameters decreases as $Q$ increases.

We introduce $u=\tfrac{1}{r}$. From Eq.~(\ref{eq8}), we can obtain
\begin{equation}\label{eq12}
\frac{d r}{d \varphi}=\pm\,r^2 \sqrt{\frac{1}{b_c^2}-\frac{1}{r^2}\left(\frac{1}{1-l}-\frac{2 M}{r}
+\frac{Q^2}{(1-l)^2\,r^2}\right)},
\end{equation}
thus
\begin{equation}\label{eq13}
\frac{d u}{d \varphi}=
\sqrt{\frac{1}{b_c^2}-u^2\left(\frac{1}{1-l}-2 M u+\frac{Q^2 u^2}{(1-l)^2}\right)}.
\end{equation}
Photons with $b_c<b_h$ fall into the black hole, those with $b_c=b_h$ perpetually orbit on the photon sphere, and those with $b_c>b_h$ escape to infinity.

\section{The image of an accretion disk around a charged black hole in the Kalb-Ramond background}\label{sec:QNMs}

\subsection{Classification of photon trajectories}\label{sec:QNMs1}
For an observer at infinity, we analyze trajectories governed by Eq.~(\ref{eq12}). By transforming, one finds
\begin{equation}
d \varphi=\frac{d u}{\sqrt{\tfrac{1}{b_c^2}-u^2\bigl(\tfrac{1}{1-l}-2 M u+\tfrac{Q^2 u^2}{(1-l)^2}\bigr)}},
\end{equation}
whose integral yields
\begin{equation}
\varphi=\int \frac{d u}{\sqrt{\tfrac{1}{b_c^2}-u^2\bigl(\tfrac{1}{1-l}-2M u+\tfrac{Q^2 u^2}{(1-l)^2}\bigr)}}.
\end{equation}
We classify photon trajectories based on the number of times $n$ they cross the equatorial plane
\begin{equation}\label{eq16}
n=\frac{\varphi}{2 \pi}.
\end{equation}
We define: 
- Direct orbits if $n<0.75$, 
- Lensing orbits if $0.75<n<1.25$, 
- Photon orbits if $n>1.25$.

\begin{table}[htbp]
\centering
\caption{The range of $b_c$ for direct, lensing, and photon orbits under different $Q$ values.}
\label{PRRR}
\begin{tabular}{|c|c|c|c|}
\hline 
$Q$ & Direct $(n<0.75)$ & Lensing $(0.75<n<1.25)$ & Photon $(n>1.25)$\\
\hline
0.1 & $b_c<4.93478$ or $b_c>6.03871$ & $4.93478<b_c<5.10164$ and $5.13981<b_c<6.03871$ & $5.10164<b_c<5.13981$ \\
\hline
0.3 & $b_c<4.85905$ or $b_c>5.97711$ & $4.85905<b_c<5.0294$ and $5.06918<b_c<5.97711$ & $5.0294<b_c<5.06918$ \\
\hline
0.6 & $b_c<4.21945$ or $b_c>5.50968$ & $4.21945<b_c<4.43842$ and $4.50112<b_c<5.50968$ & $4.43842<b_c<4.50112$ \\
\hline
\end{tabular}
\end{table}

\begin{figure*}[htbp]
	\centering
	\subfigure[$Q=0.1$]{
		\begin{minipage}[t]{0.28\linewidth}
			\centering
			\includegraphics[width=2.1in]{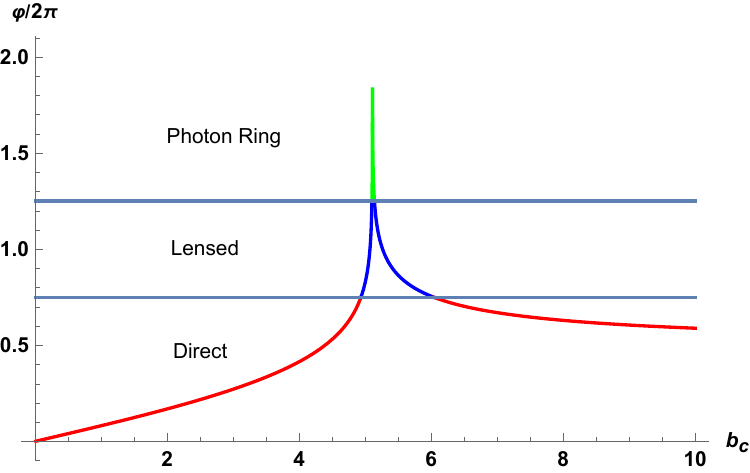}\\      
            \end{minipage}
	}
	\subfigure[$Q=0.3$]{
		\begin{minipage}[t]{0.28\linewidth}
			\centering
			\includegraphics[width=2.1in]{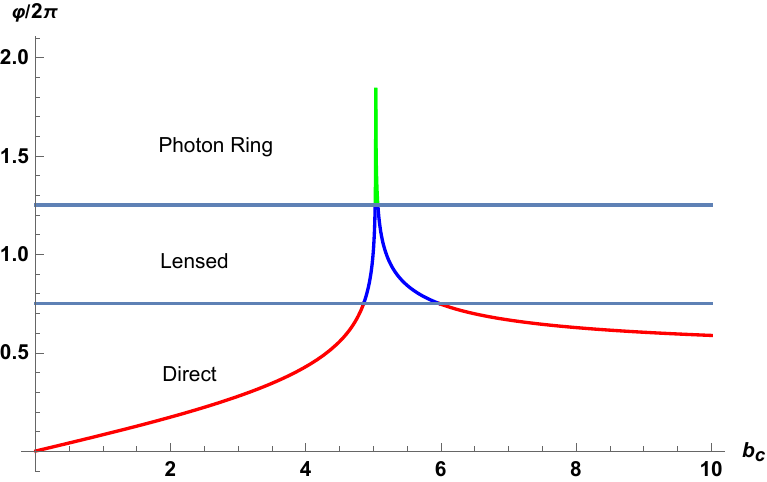}\\         
            \end{minipage}
	}
	\subfigure[$Q=0.6$]{
		\begin{minipage}[t]{0.28\linewidth}
			\centering
			\includegraphics[width=2.1in]{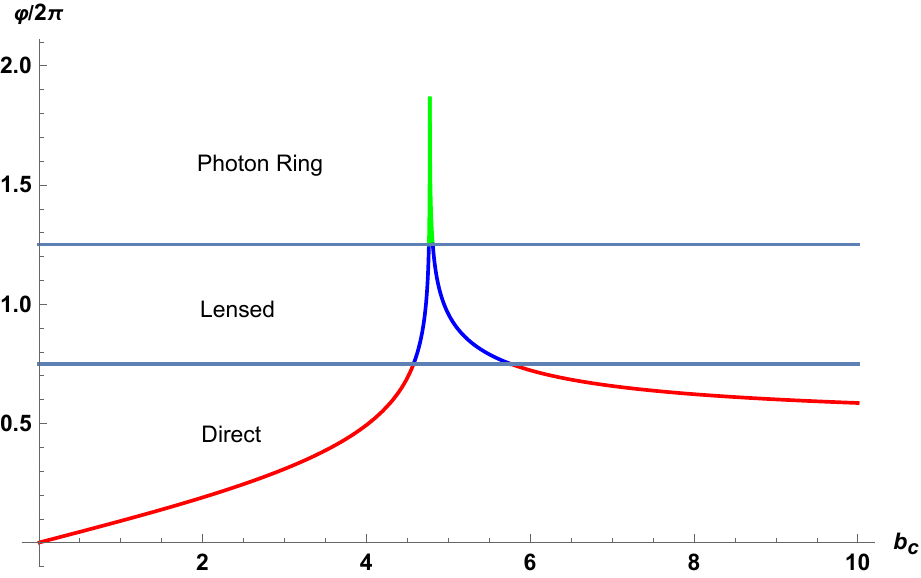}\\          \end{minipage}
	}
	\caption{(color online) Variation of intersection number $n$ with impact parameter $b_c$ for $M=1$ and different $Q$. Red curves represent direct orbits $(n<0.75)$, blue curves lensing orbits $(0.75<n<1.25)$, and green curves photon orbits $(n>1.25)$.}
	\label{fig.TF}
\end{figure*}

\begin{figure*}[htbp]
	\centering
	\subfigure[$Q=0.1$]{
		\begin{minipage}[t]{0.28\linewidth}
			\centering
			\includegraphics[width=2.1in]{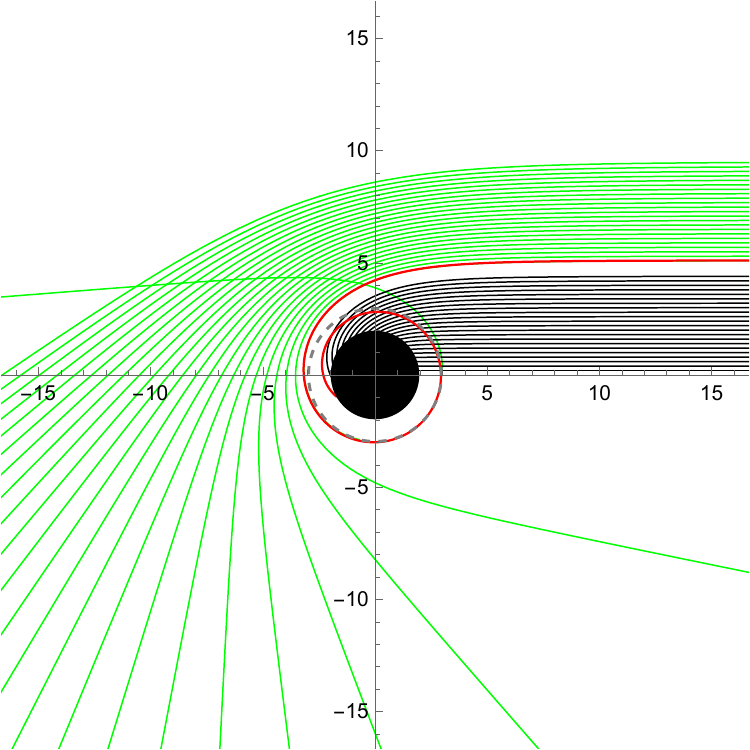}\\
		\end{minipage}
	}
	\subfigure[$Q=0.3$]{
		\begin{minipage}[t]{0.28\linewidth}
			\centering
			\includegraphics[width=2.1in]{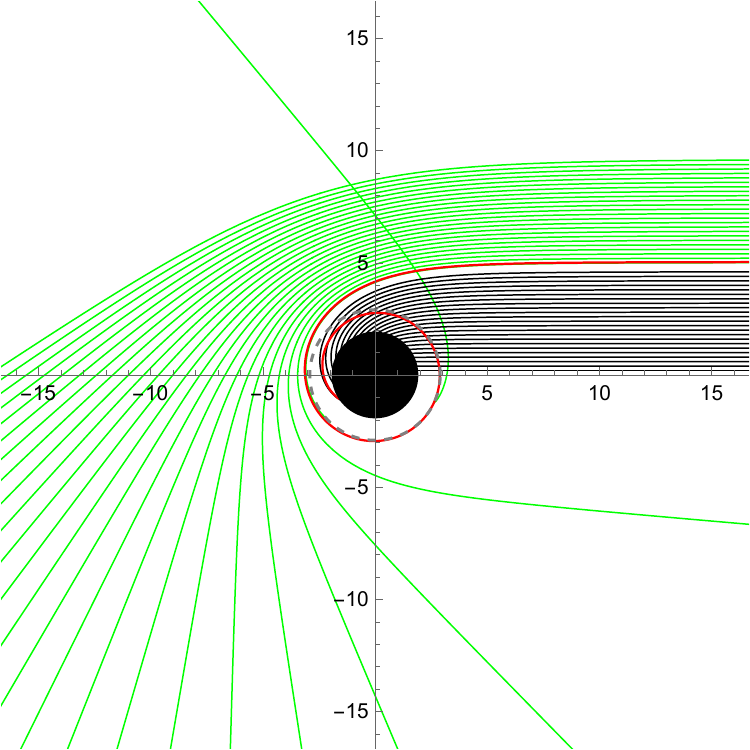}\\
		\end{minipage}
	}
	\subfigure[$Q=0.6$]{
		\begin{minipage}[t]{0.28\linewidth}
			\centering
			\includegraphics[width=2.1in]{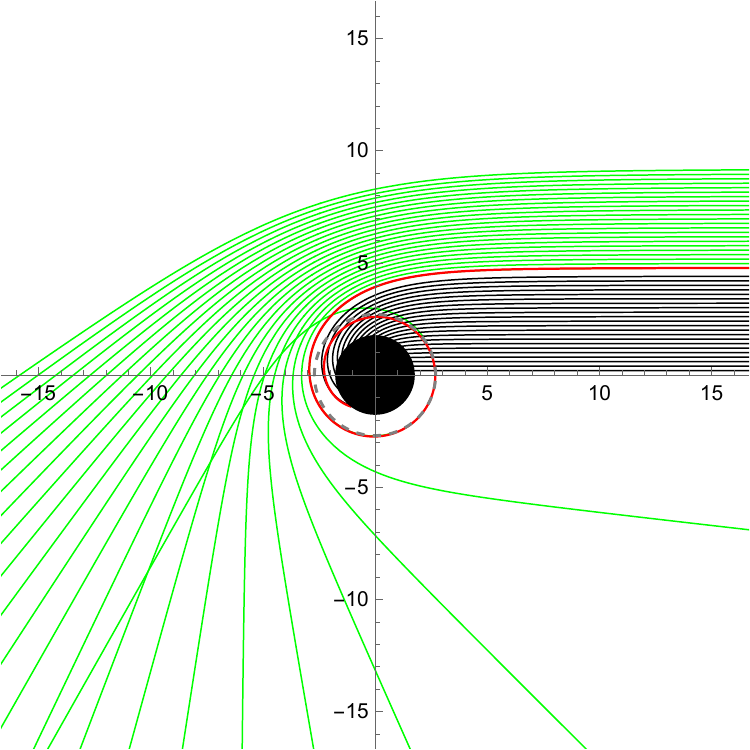}\\
		\end{minipage}
	}
	\caption{(color online) Polar-coordinate trajectories of photons for $M=1$ under different $Q$. The black disk denotes the black hole horizon, and gray dashed circles mark the photon spheres.}
	\label{fig.TFF}
\end{figure*}

\begin{figure*}[htbp]
	\centering
	\subfigure[$Q=0.1$]{
		\begin{minipage}[t]{0.28\linewidth}
			\centering
			\includegraphics[width=2.1in]{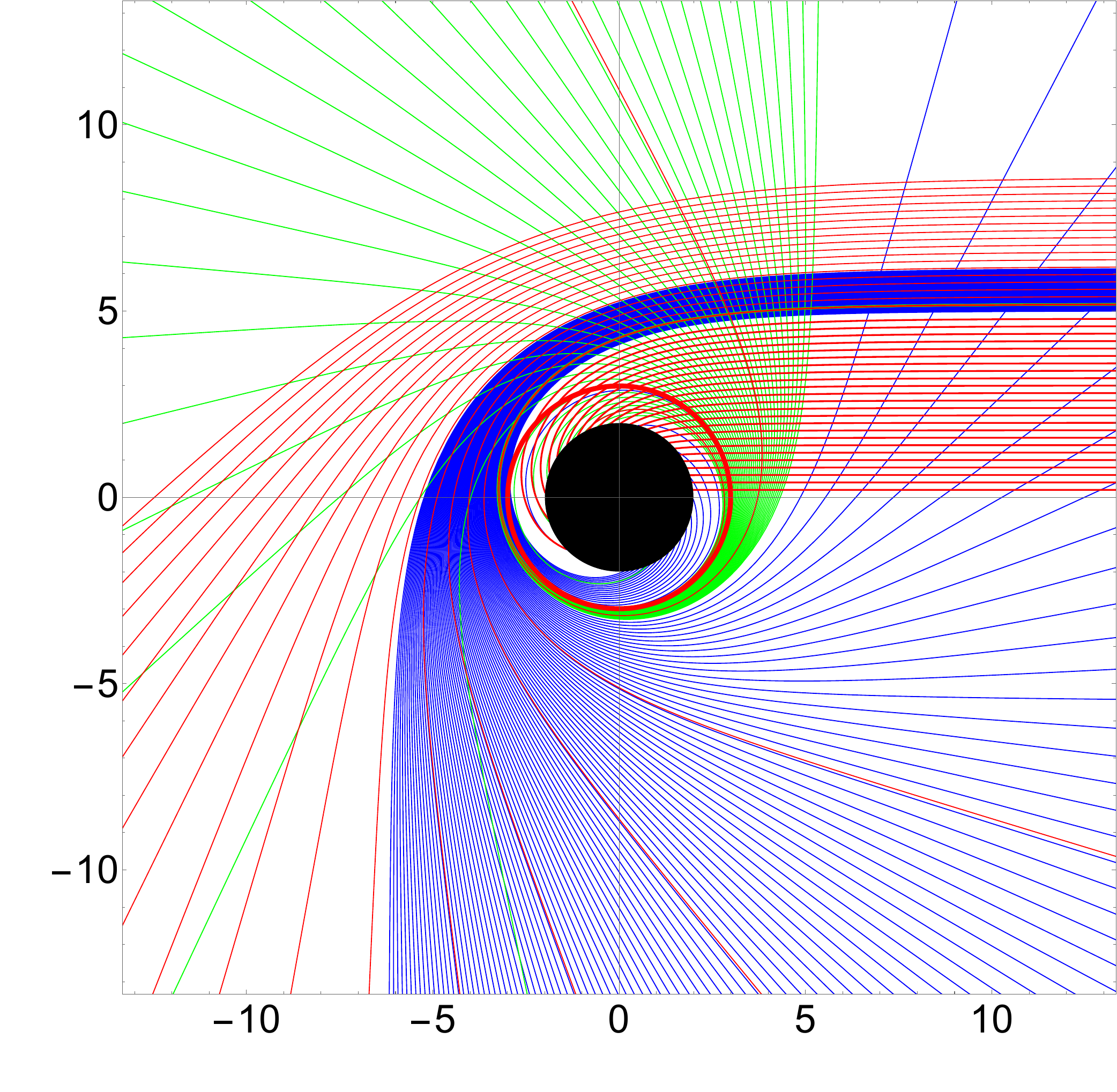}\\
		\end{minipage}
	}
	\subfigure[$Q=0.3$]{
		\begin{minipage}[t]{0.28\linewidth}
			\centering
			\includegraphics[width=2.1in]{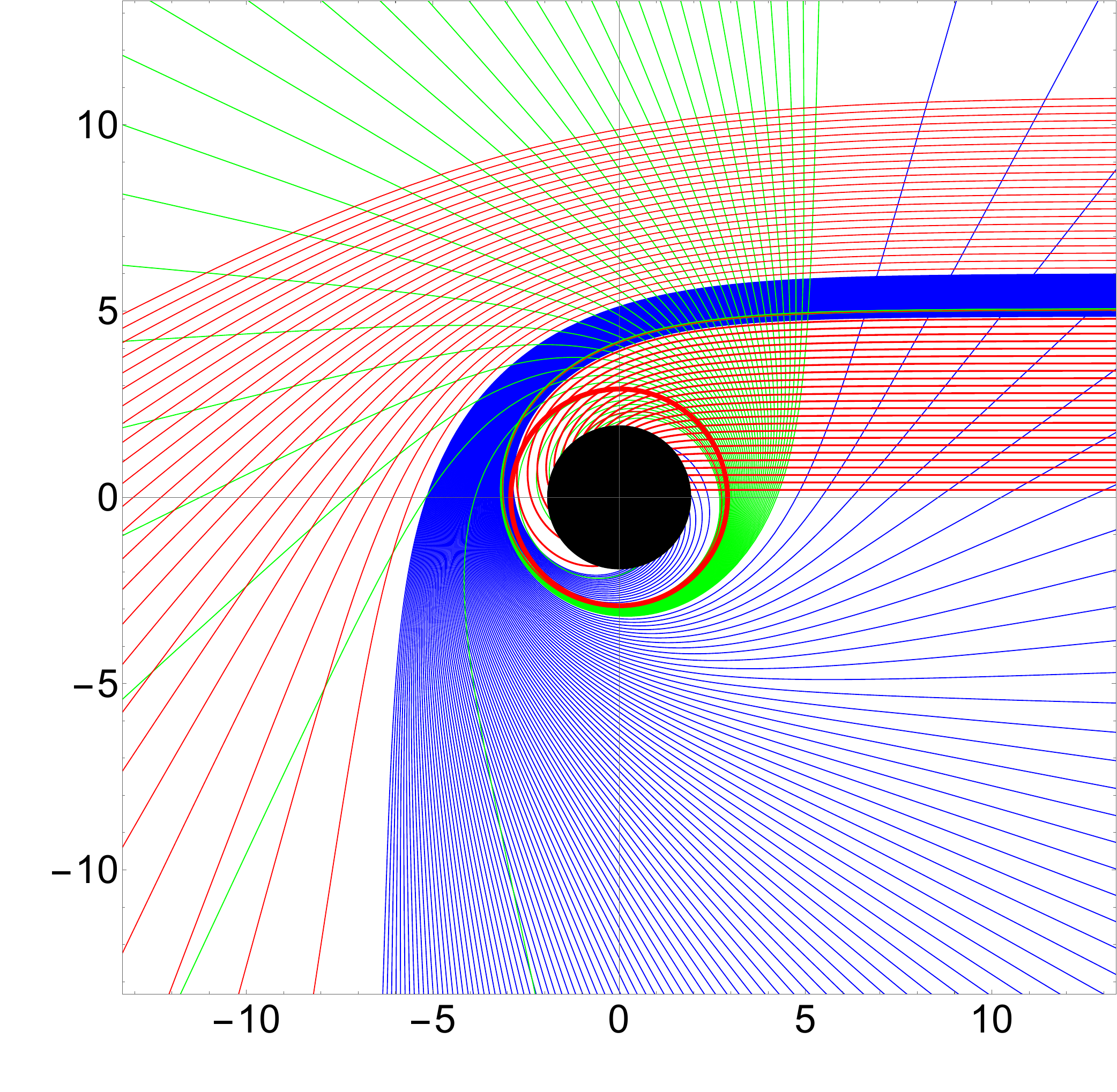}\\
		\end{minipage}
	}
	\subfigure[$Q=0.6$]{
		\begin{minipage}[t]{0.28\linewidth}
			\centering
			\includegraphics[width=2.1in]{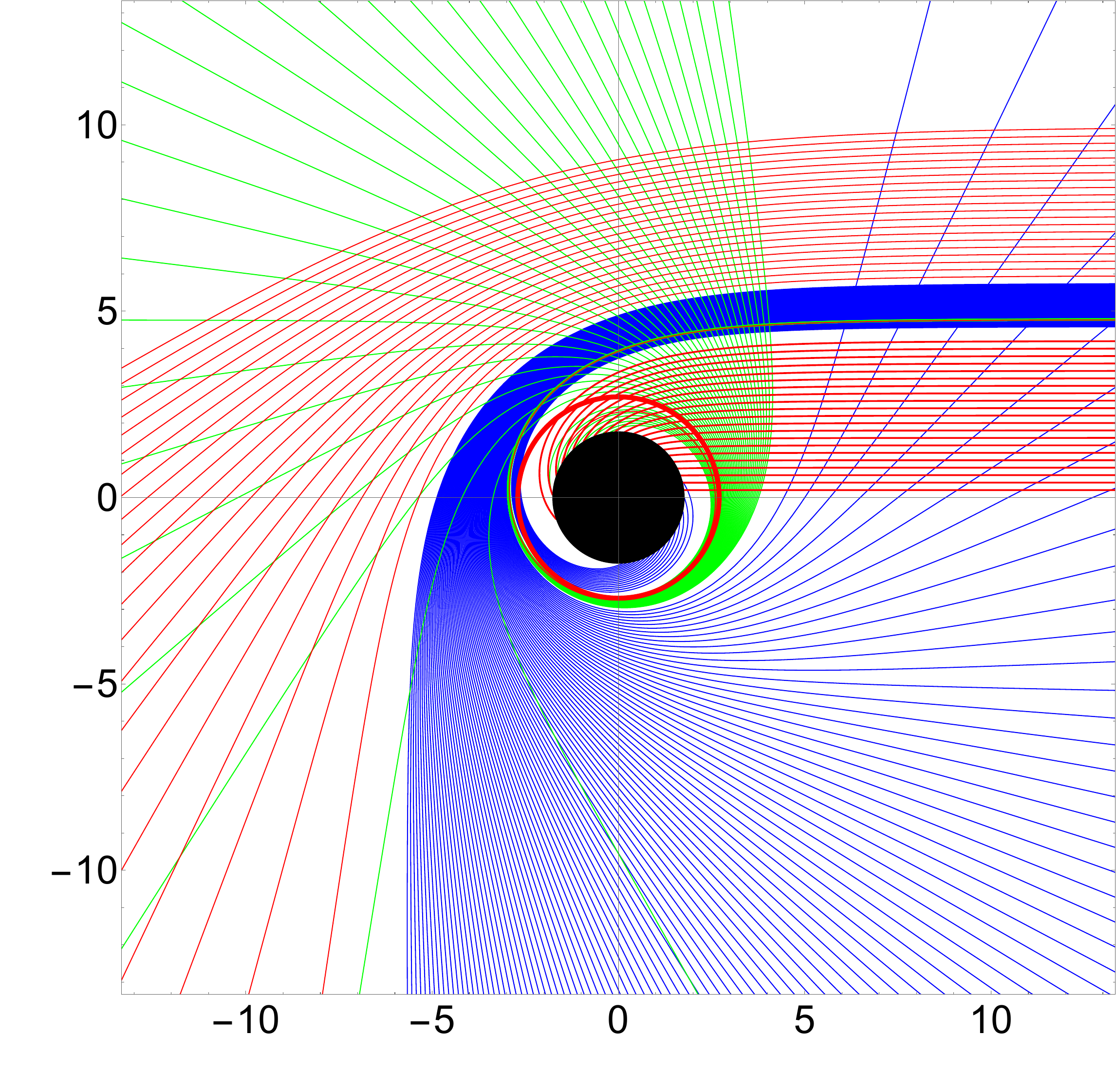}\\
		\end{minipage}
	}
	\caption{(color online) Photon trajectories for $M=1$ and different $Q$. The black region represents the black hole horizon, the red circle is the photon sphere, and red/blue/green lines correspond to direct, lensing, and photon orbits, respectively.}
	\label{fig.TFFF}
\end{figure*}

Fig.\ref{fig.TF} directly shows the classification of trajectories. As $Q$ increases from 0.1 to 0.6, the relevant ranges of $b_c$ for direct/lensing/photon orbits shrink, and the radius of the photon sphere decreases. Fig.\ref{fig.TFFF} further illustrates the photon orbits in polar coordinates, showing how larger $Q$ decreases both the photon sphere radii and horizon radius but slightly increases their thickness.

\subsection{The black hole transfer function with $l=0.01$}
Consider a thin accretion disk lying in the equatorial plane, observed from the north pole at infinity. The emission is assumed isotropic. The observed intensity is proportional to the number of disk-crossings. By Liouville’s theorem~\cite{Gralla:2019xty}
\begin{equation}
I_\nu^{\mathrm{obs}}(r)=F(r)^{\frac{3}{2}}\; I_\nu^{\mathrm{em}}(r).
\end{equation}
Integrating over frequency gives
\begin{equation}
I_1(r)=\int I_\nu^{\mathrm{obs}}(r)\,d\nu =F(r)^2 I_\nu^{\mathrm{em}}(r).
\end{equation}
Summing over each intersection $m$ with the disk yields
\begin{equation}
I_1(r)=\sum_m F(r)^2\,I_\nu^{\mathrm{em}}\bigl|_{r=r_m(b_c)},
\end{equation}
where $r_m(b_c)$ is the transfer function identifying the radial position on the disk for each crossing.

\begin{figure*}[htbp]
	\centering
	\subfigure[$Q=0.1$]{
		\begin{minipage}[t]{0.28\linewidth}
			\centering
			\includegraphics[width=2.1in]{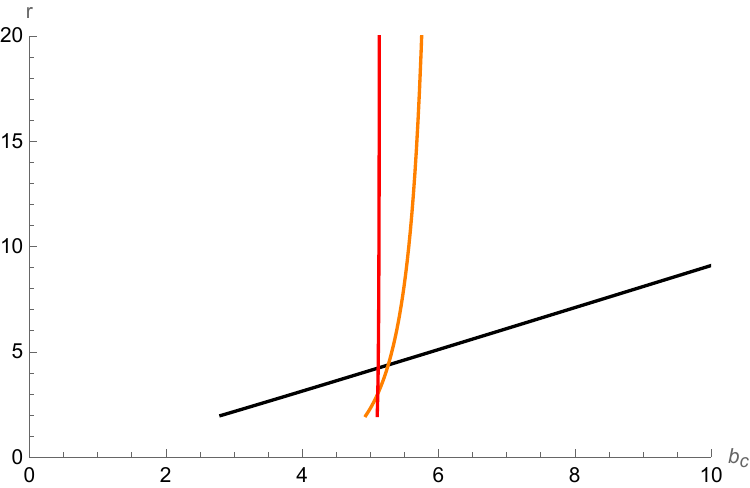}\\
		\end{minipage}
	}
	\subfigure[$Q=0.3$]{
		\begin{minipage}[t]{0.28\linewidth}
			\centering
			\includegraphics[width=2.1in]{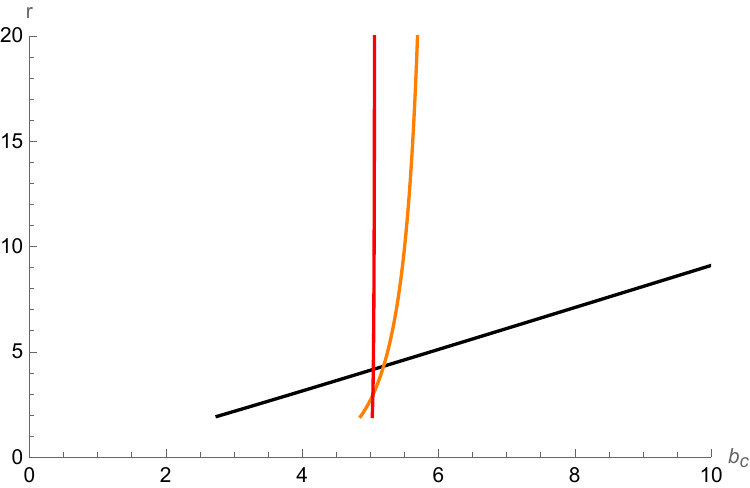}\\
		\end{minipage}
	}
	\subfigure[$Q=0.6$]{
		\begin{minipage}[t]{0.28\linewidth}
			\centering
			\includegraphics[width=2.1in]{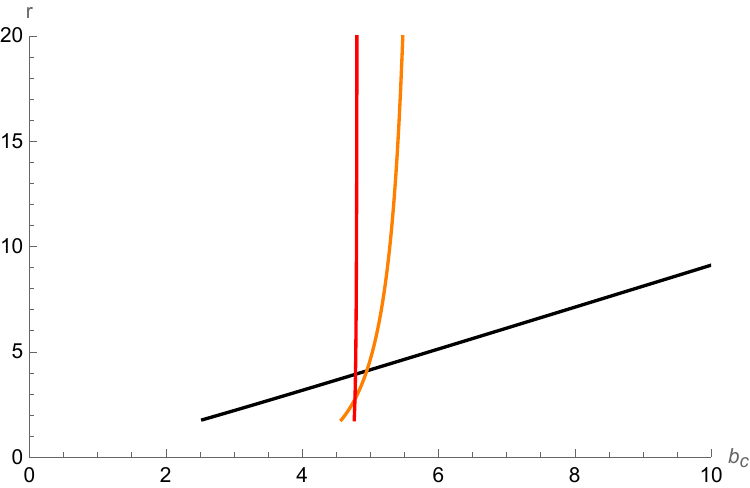}\\
		\end{minipage}
	}
	\caption{(color online) The transfer function for different $Q$ values. Black curves correspond to direct rings, orange to lensing rings, and red to photon rings.}
	\label{fig.TFFFF}
\end{figure*}

Fig.\ref{fig.TFFFF} plots the first three transfer functions. The slope of each curve is its demagnification factor. Direct emission ($m=1$) has a near-constant slope. Lensing ($m=2$) has a steeper slope, and the photon ring ($m=3$) shows an almost infinite slope. Hence, direct emission dominates the total observed intensity, whereas lensing and photon rings are strongly demagnified. As $Q$ increases, all three curves shift to smaller radii. This analytical result is consistent with our numerical simulations, further supporting our conclusion.

\subsection{The optical appearance of a charged black hole surrounded by a thin accretion disk}
We consider three emission models:

1. \textbf{Model A}: Radiation starts from the innermost stable circular orbit ($r_{\mathrm{ISCO}}$) with a quadratic decay~\cite{He:2022yse}
   \begin{equation}
   I_a^{\mathrm{em}}(r)=
   \begin{cases}
   \left(\tfrac{1}{r-(r_{\mathrm{ISCO}}-1)}\right)^2, & r \geq r_{\mathrm{ISCO}},\\
   0, & r<r_{\mathrm{ISCO}}.
   \end{cases}
   \end{equation}
   The ISCO in spherical symmetry is
   \[
   r_{\mathrm{ISCO}}
   =\frac{3\,F(r_{\mathrm{ISCO}})\,F'(r_{\mathrm{ISCO}})}{2\,F'(r_{\mathrm{ISCO}})-F(r_{\mathrm{ISCO}})\,F''(r_{\mathrm{ISCO}})}.
   \]

2. \textbf{Model B}: Radiation starts from the photon sphere $r_p$ and decays cubically~\cite{He:2022yse}
   \begin{equation}
   I_b^{\mathrm{em}}(r)=
   \begin{cases}
   \left(\tfrac{1}{r-(r_p-1)}\right)^3, & r \geq r_p,\\
   0, & r<r_p.
   \end{cases}
   \end{equation}

3. \textbf{Model C}: Radiation originates from the event horizon $r_h$ with a certain arctangent-based decay~\cite{He:2022yse}
   \begin{equation}
   I_c^{\mathrm{em}}(r)=
   \begin{cases}
   \dfrac{\frac{\pi}{2}-\tan ^{-1}\bigl(r-(r_{\mathrm{ISCO}}-1)\bigr)}
   {\frac{\pi}{2}-\tan ^{-1}\bigl(r_h\bigr)}, & r \geq r_h,\\
   0, & r<r_h.
   \end{cases}
   \end{equation}

\begin{figure*}[htbp]
	\centering
	\subfigure[$Q=0.1$]{
		\begin{minipage}[t]{0.28\linewidth}
			\includegraphics[width=2.1in]{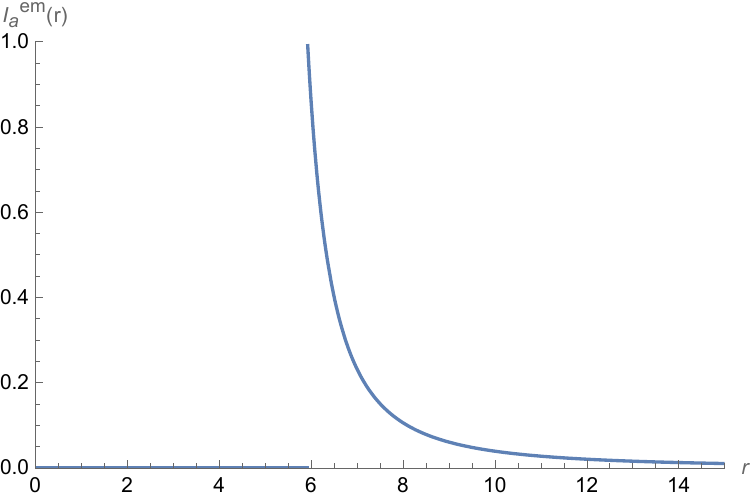}\\
			\includegraphics[width=2.1in]{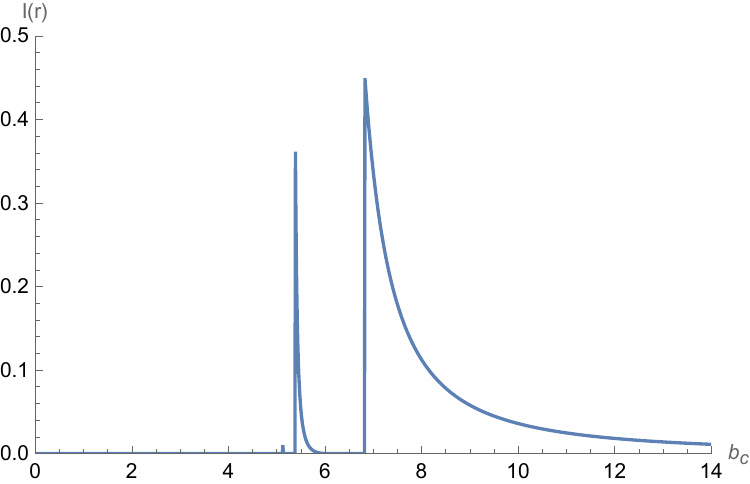}\\
			\includegraphics[width=2.1in]{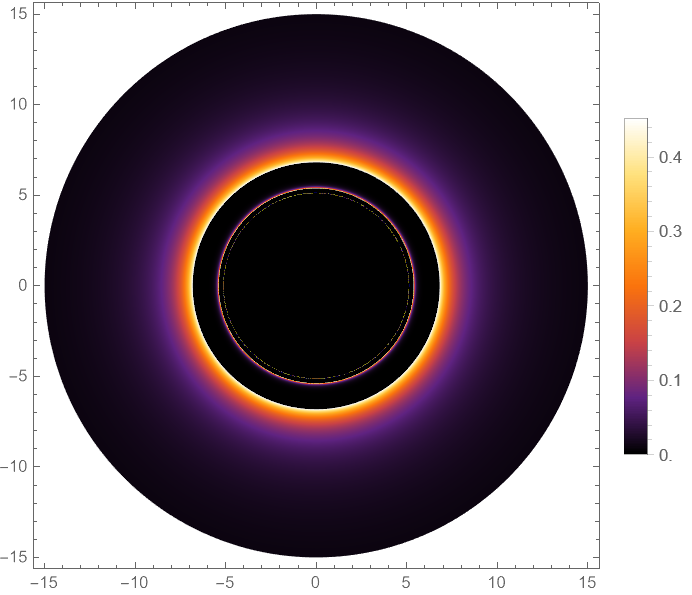}\\
		\end{minipage}
	}
	\subfigure[$Q=0.3$]{
		\begin{minipage}[t]{0.28\linewidth}
			\includegraphics[width=2.1in]{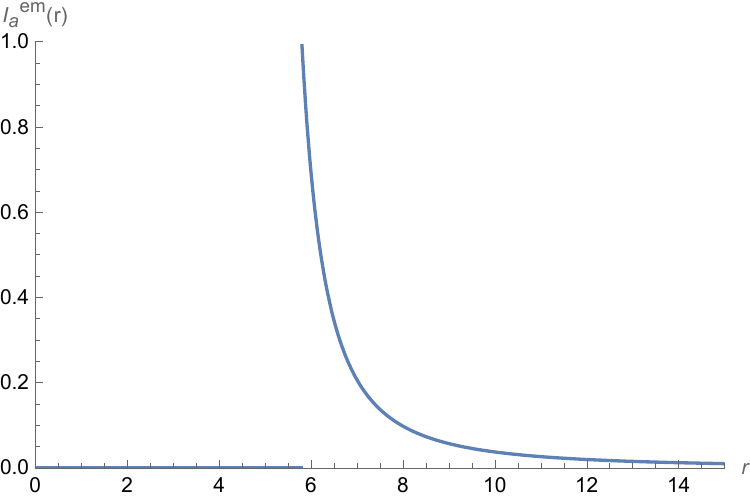}\\
			\includegraphics[width=2.1in]{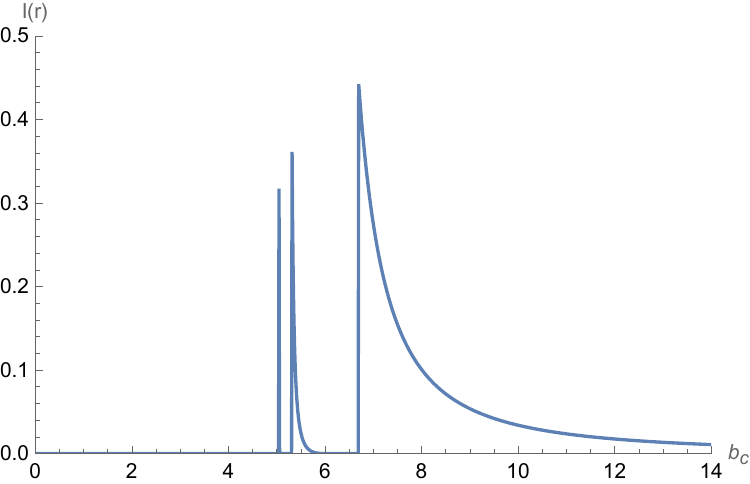}\\
			\includegraphics[width=2.1in]{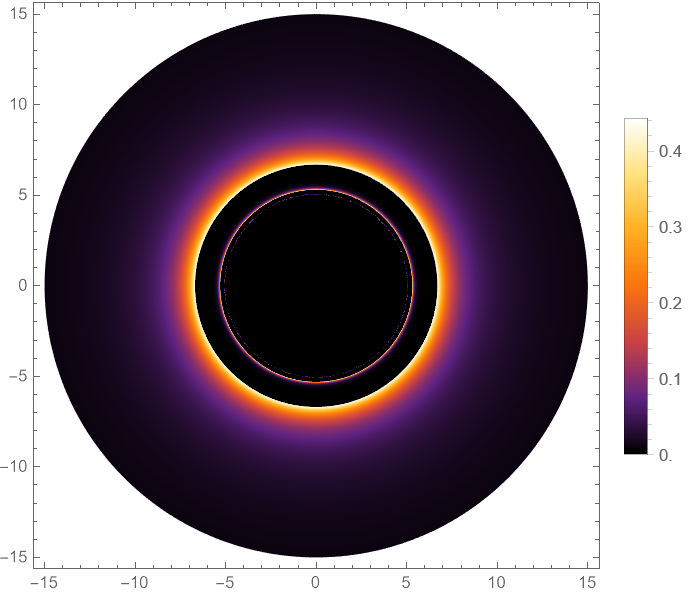}\\
		\end{minipage}
	}
	\subfigure[$Q=0.6$]{
		\begin{minipage}[t]{0.28\linewidth}
			\includegraphics[width=2.1in]{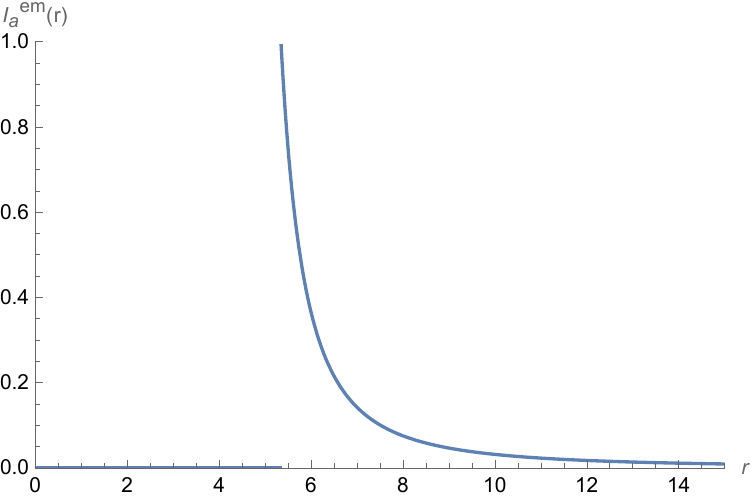}\\
			\includegraphics[width=2.1in]{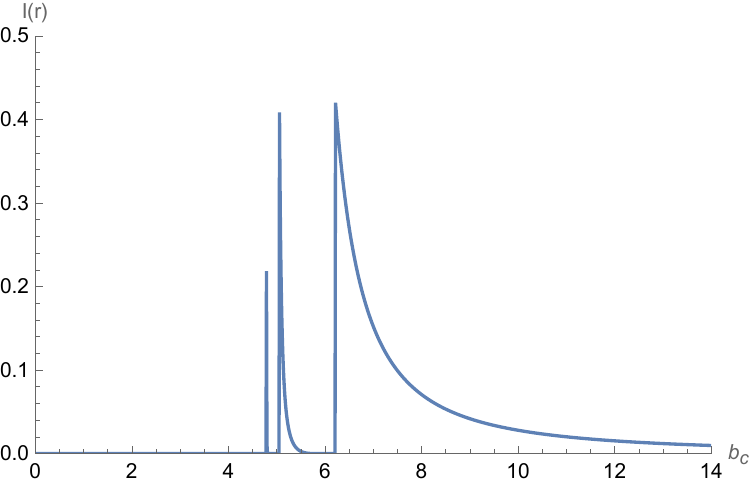}\\
			\includegraphics[width=2.1in]{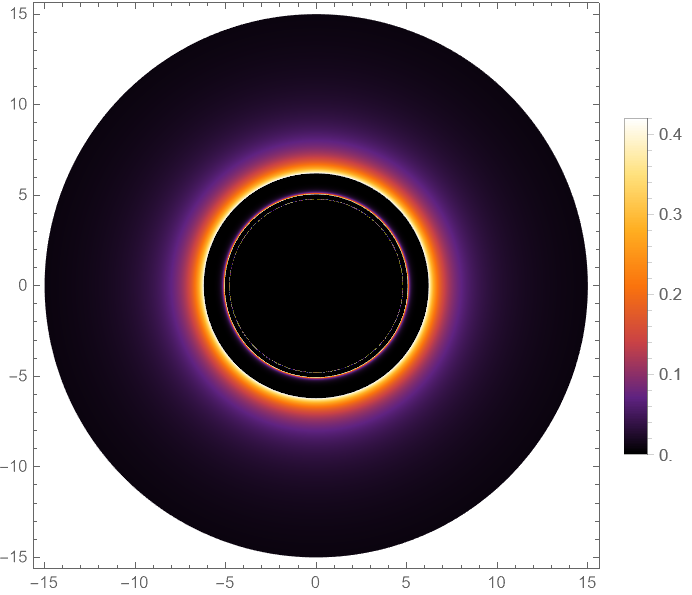}\\
		\end{minipage}
	}
	\caption{(color online) Model A results, where radiation starts from $r_{\mathrm{ISCO}}$. The first row: emission intensity vs.\ $r$. The second row: observed intensity vs.\ $b_c$. The third row: two-dimensional appearance in the celestial plane.}
	\label{fig.IS}
\end{figure*}

\begin{figure*}[htbp]
	\centering
	\subfigure[$Q=0.1$]{
		\begin{minipage}[t]{0.28\linewidth}
			\includegraphics[width=2.1in]{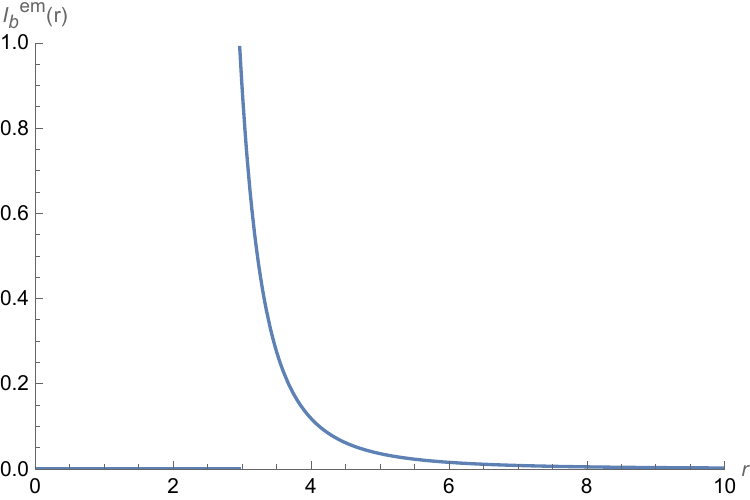}\\
			\includegraphics[width=2.1in]{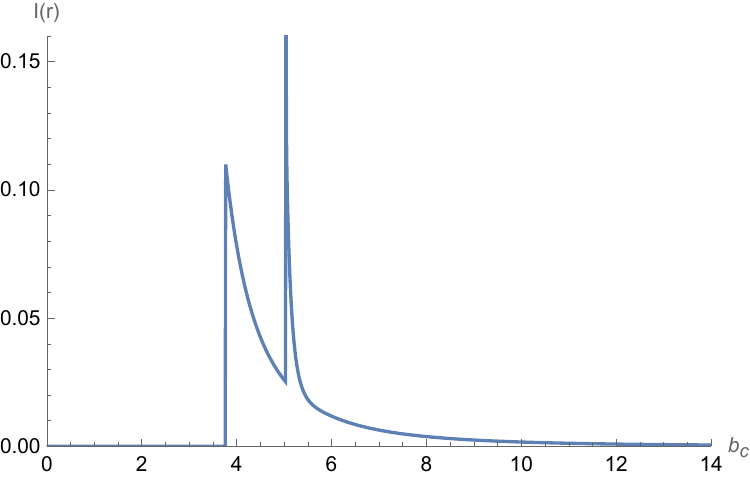}\\
			\includegraphics[width=2.1in]{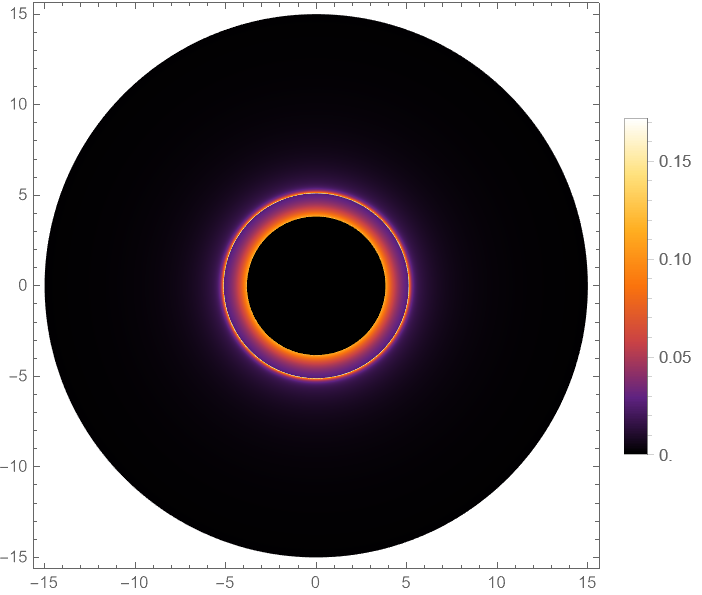}\\
		\end{minipage}
	}
	\subfigure[$Q=0.3$]{
		\begin{minipage}[t]{0.28\linewidth}
			\includegraphics[width=2.1in]{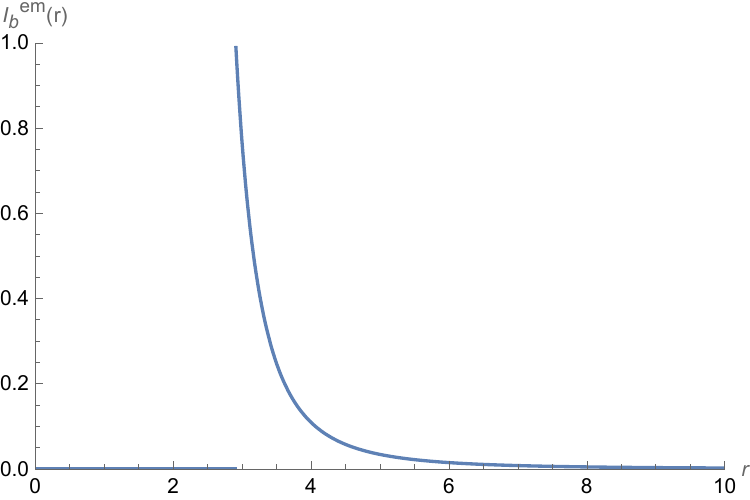}\\
			\includegraphics[width=2.1in]{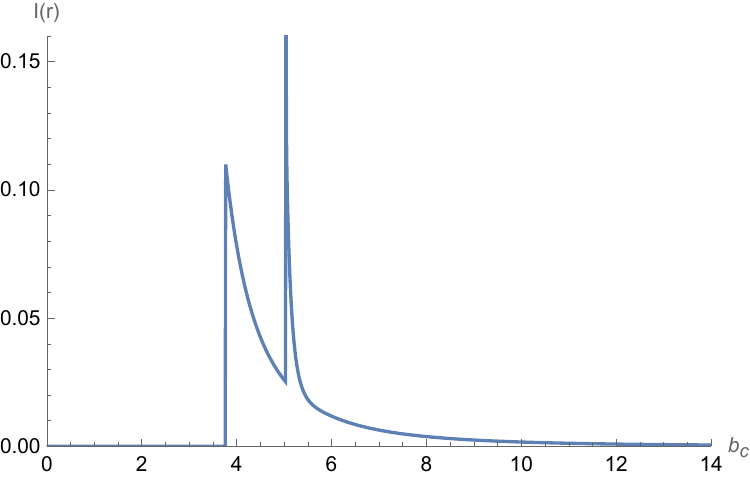}\\
			\includegraphics[width=2.1in]{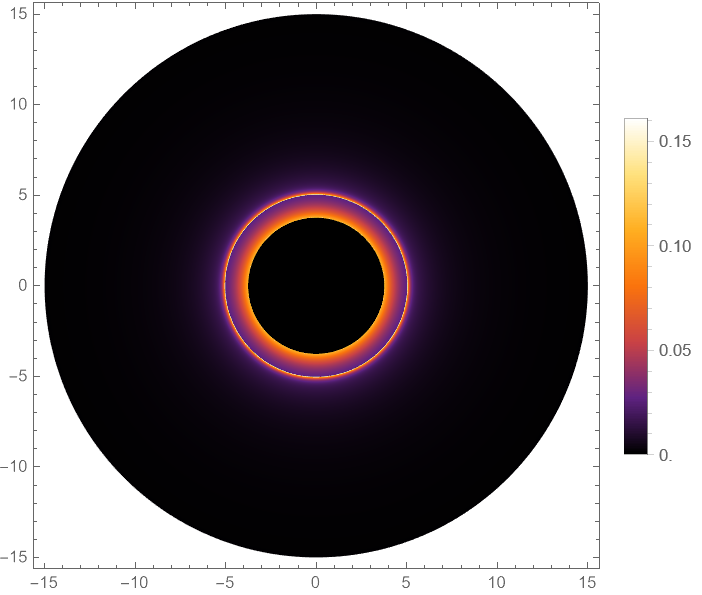}\\
		\end{minipage}
	}
	\subfigure[$Q=0.6$]{
		\begin{minipage}[t]{0.28\linewidth}
			\includegraphics[width=2.1in]{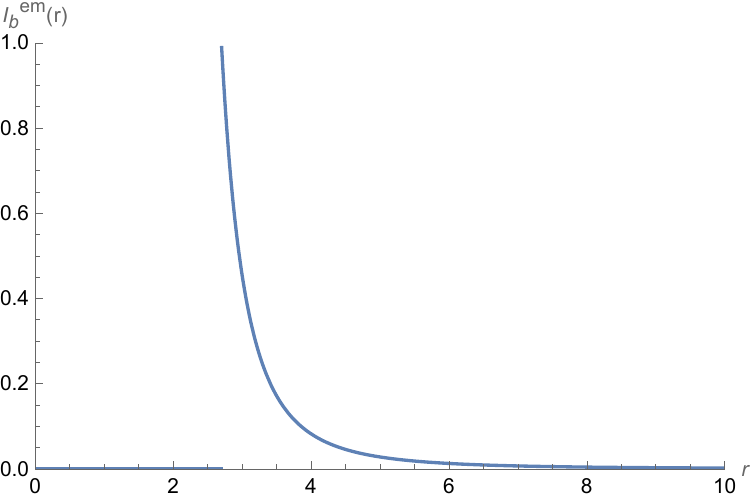}\\
			\includegraphics[width=2.1in]{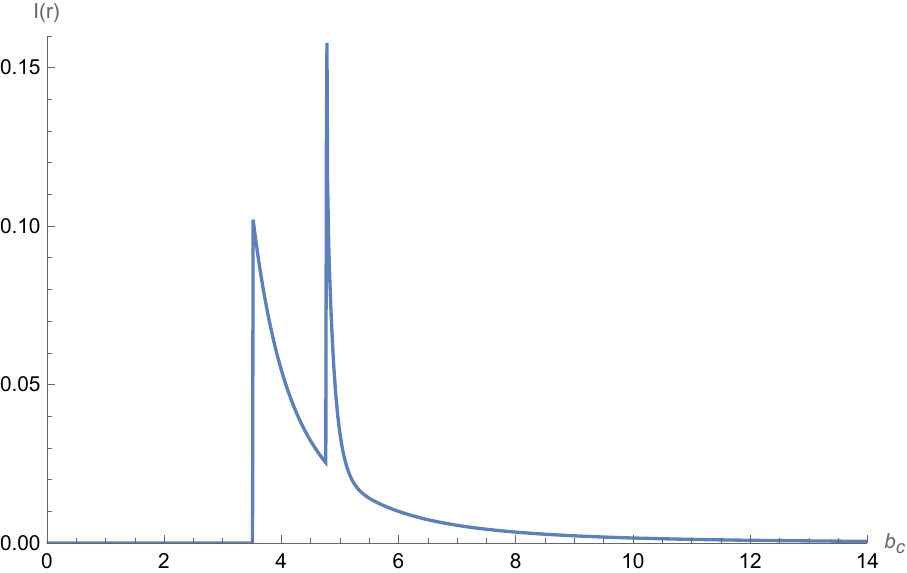}\\
			\includegraphics[width=2.1in]{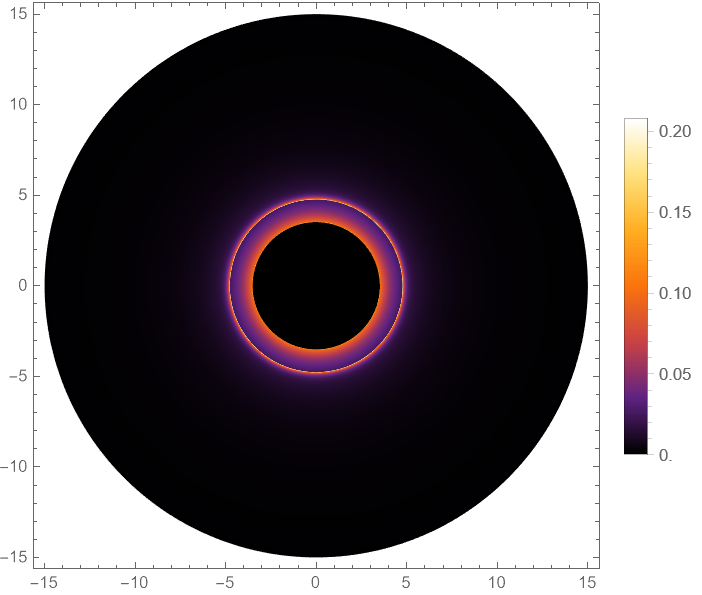}\\
		\end{minipage}
	}
	\caption{(color online) Model B results, where radiation starts from $r_p$. Same plotting scheme as Fig.~\ref{fig.IS}.}
	\label{fig.ISS}
\end{figure*}

\begin{figure*}[htbp]
	\centering
	\subfigure[$Q=0.1$]{
		\begin{minipage}[t]{0.28\linewidth}
			\includegraphics[width=2.1in]{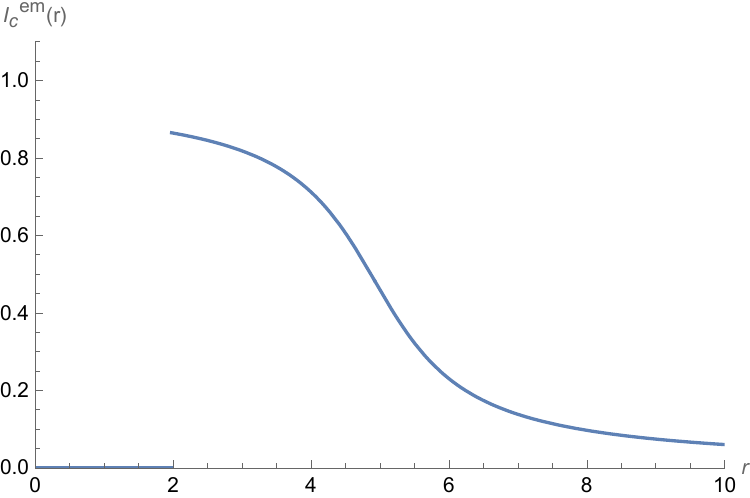}\\
			\includegraphics[width=2.1in]{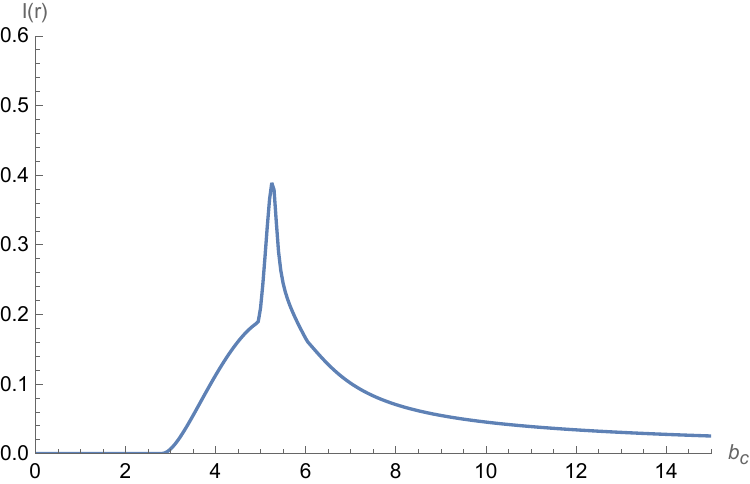}\\
			\includegraphics[width=2.1in]{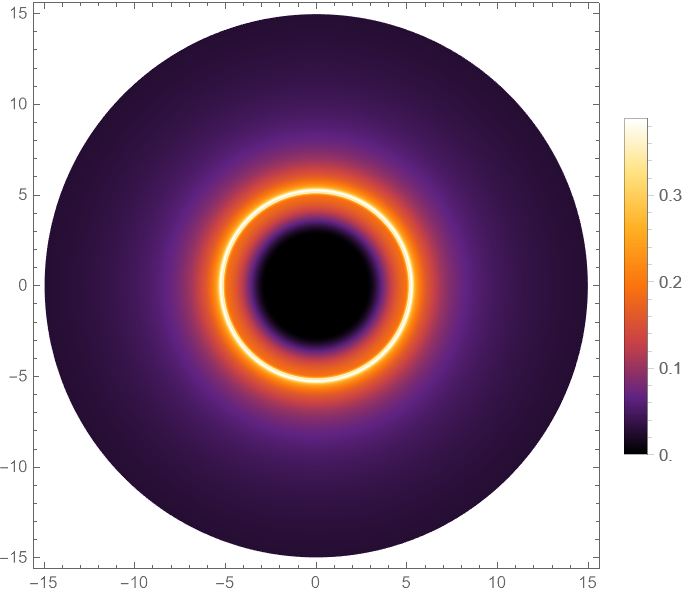}\\
		\end{minipage}
	}
	\subfigure[$Q=0.3$]{
		\begin{minipage}[t]{0.28\linewidth}
			\includegraphics[width=2.1in]{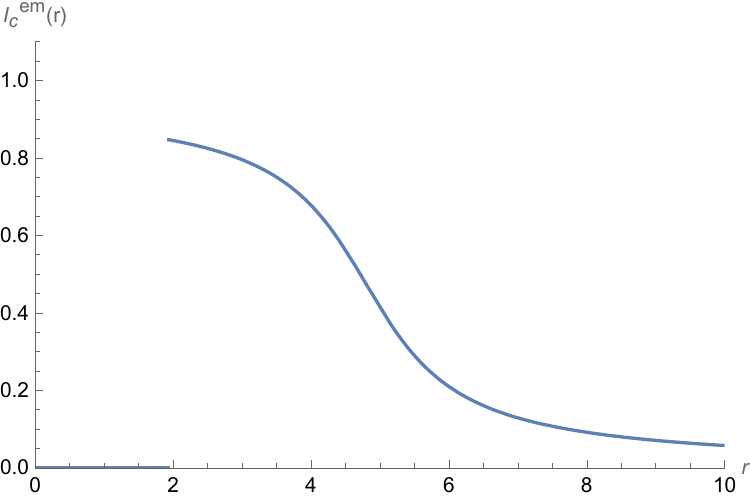}\\
			\includegraphics[width=2.1in]{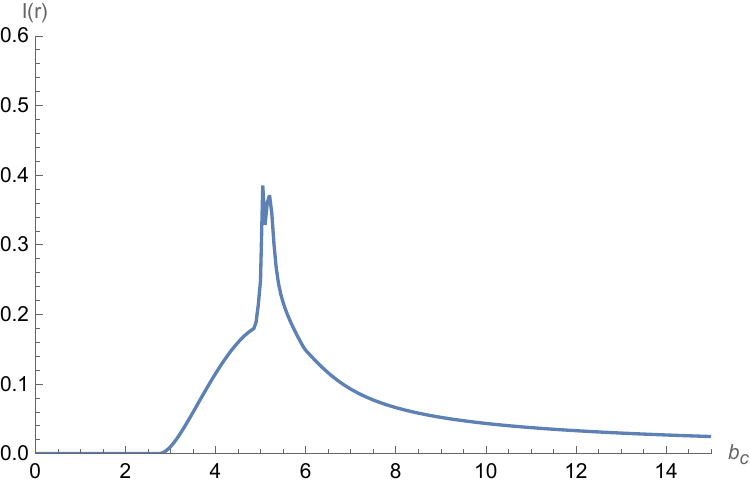}\\
			\includegraphics[width=2.1in]{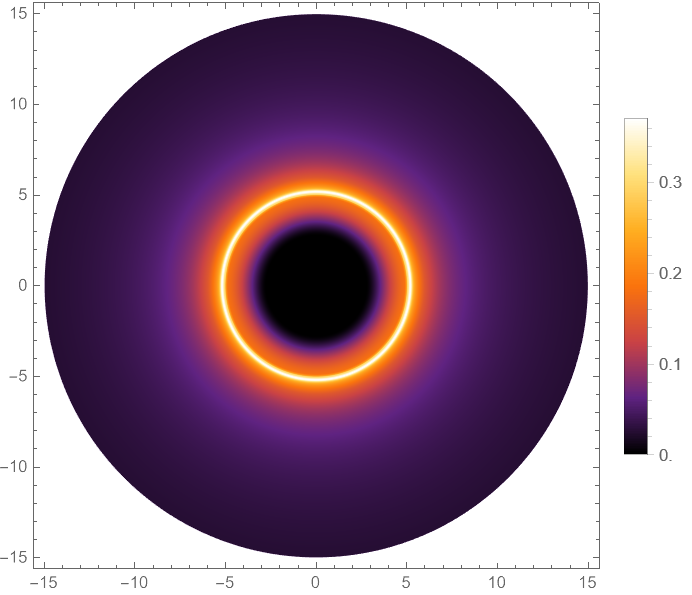}\\
		\end{minipage}
	}
	\subfigure[$Q=0.6$]{
		\begin{minipage}[t]{0.28\linewidth}
			\includegraphics[width=2.1in]{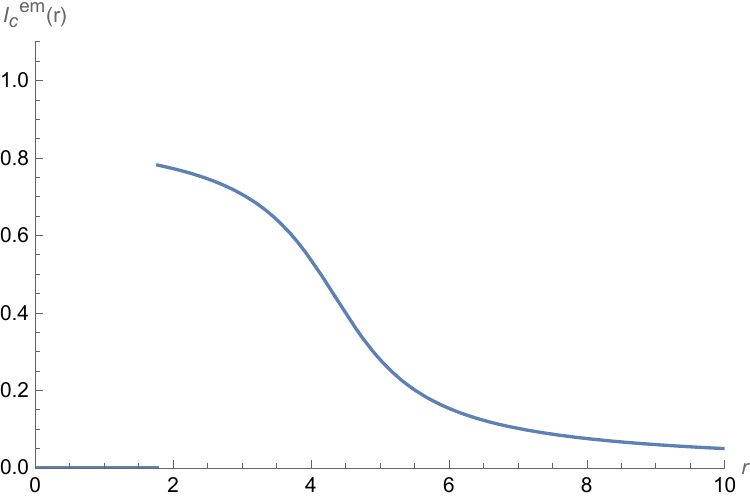}\\
			\includegraphics[width=2.1in]{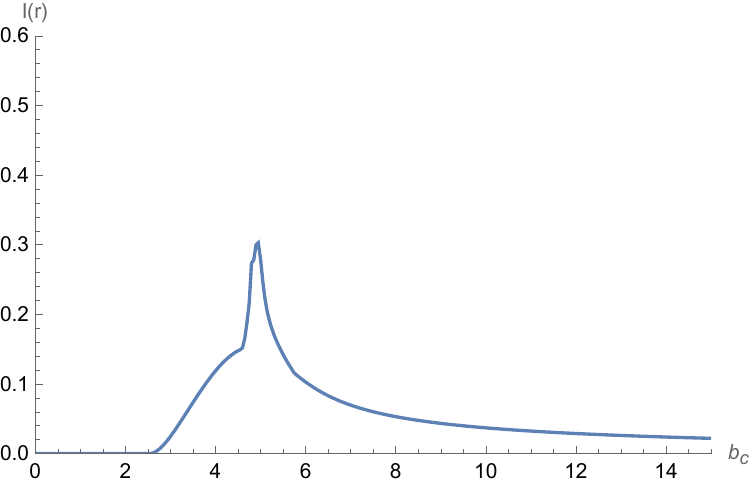}\\
			\includegraphics[width=2.1in]{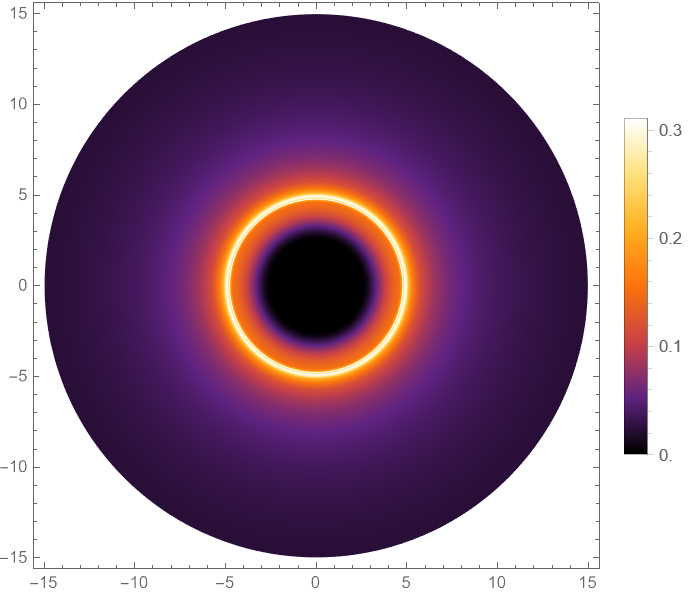}\\
		\end{minipage}
	}
	\caption{(color online) Model C results, where radiation starts from $r_h$. Same plotting scheme as Fig.~\ref{fig.IS}.}
	\label{fig.ISSS}
\end{figure*}

Figs. \ref{fig.IS}, \ref{fig.ISS}, and \ref{fig.ISSS} show the emission intensity vs.\ $r$, observed intensity vs.\ $b_c$, and two-dimensional snapshots for Models A, B, and C, respectively, at different $Q$ values.

Using the first model, we have constructed Fig. \ref{fig.IS}. The first row of Fig. \ref{fig.IS} represents the relationship between the emission intensity and $r$. It can be observed that the emission intensity reaches a peak when $r=r_{I S C O}$ and then rapidly decays to zero. When the charge quantity $Q=0.1$, the emission peak is at $r \approx 5.92 \mathrm{M}$. When the charge quantity $Q=0.3$, the emission peak is at $r \approx 5.78 \mathrm{M}$. When the charge quantity $Q=0.6$, the emission peak is at $r \approx 5.32 \mathrm{M}$. We find that the larger the charge, the smaller the position of the curve peak, and the peak value also slightly decreases. 

This trend can be explained by the increasing gravitational influence of the charged black hole, which causes the emission region to contract. The decreased emission intensity with increasing $Q$ also suggests a possible modification in the underlying emission mechanisms as the charge affects the photon trajectories.

The second row represents the relationship between the observed intensity by the observer and the impact parameter $b_c$. It can be observed that the three images in the second row all have three peaks. These three peaks correspond to the photon sphere, lens ring, and direct ring as $b_c$ increases. When the charge quantity $Q=0.1$, the observed intensity peaks occur at $b_c \approx 5.12 \mathrm{M}$, $b_c \approx 5.36 \mathrm{M}$, and $b_c \approx 6.80 \mathrm{M}$. Similarly, when $Q=0.3$, the peaks of the observed intensities are located at $b_c \approx 5.01 \mathrm{M}$, $b_c \approx 5.30 \mathrm{M}$, and $b_c \approx 6.67 \mathrm{M}$. For $Q=0.6$, the observed intensity peaks are also found at $b_c \approx 4.78 \mathrm{M}$, $b_c \approx 5.04 \mathrm{M}$, and $b_c \approx 6.20 \mathrm{M}$. It can also be observed that as the charge quantity $Q$ increases, both the position of the emission peak and the position of the observed intensity peak shift inward.

In the region of the impact parameter $b_c$ corresponding to these three rings, the photon sphere is the smallest, the lens ring is smaller, and the direct ring is the largest. Therefore, the direct ring contributes the most to the observed intensity, the photon sphere contributes the least, and the lens ring contributes slightly more than the photon sphere. Thus, the observed intensity is mainly derived from direct emission. At the same charge quantity $Q$, it can be found that the position of the emission peak is smaller than the observed peak position of the direct ring, a phenomenon caused by the gravitational lensing effect. The third row represents the appearance of a two-dimensional thin disk with different charge quantities $Q$ in the celestial coordinate system. It can be discovered that each two-dimensional image has three rings, corresponding to the photon sphere, lens ring, and direct ring from the inside to the outside. Since the photon sphere contributes very little to the observed intensity, it forms a weak ring. The brightness of the lens ring is slightly greater than that of the photon sphere. The outermost direct ring is the brightest and widest because it is the main source of the total observed intensity.

Observing Fig. \ref{fig.ISS}, we have plotted the emission intensity versus radius $r$ for the second emission model, starting from the photon sphere, the intensity observed by the observer versus the impact parameter $b_c$, and the two-dimensional image of the black hole starting from the photon sphere. From the emission intensity versus radius r graph, we know that there is no radiation at point $r<r_p$. At point $r=r_p$, the radiation intensity reaches a peak. At point $r>r_p$, the radiation intensity begins to decrease. When the charge quantity $Q=0.1$, the emission peak is at $r \approx 2.96 \mathrm{M}$. When the charge quantity $Q=0.3$, the emission peak is at $r \approx 2.90 \mathrm{M}$. When the charge quantity $Q=0.6$, the emission peak is at $r \approx 2.70 \mathrm{M}$. It can be observed that the first and second emission models follow the same pattern: the larger the charge quantity, the smaller the position of the peak of the emission intensity. The peak also slightly decreases. The observed intensity of the second model differs from the first model, as the second model starts radiating from the photon sphere. It is found that the peaks of the photon sphere and the lens ring overlap, making them difficult to distinguish. The contribution of the lens ring to the observed intensity in the second model is greater than that in the first model, but similarly, the contribution of the photon sphere is the smallest, and the observed intensity still primarily comes from direct emission. Examining the two-dimensional black hole images in the third row, it can also be observed that the radiation regions of the photon sphere and the lens ring overlap. As the charge quantity $Q$ increases, the range of the central area narrows, and the brightness of the rings also diminishes. With the increase of the charge quantity $Q$, the position and peak of the emission intensity, as well as the ranges of the photon sphere, lens ring, and direct emission ring corresponding to the impact parameter $b_c$, and the peaks of the observed intensity, all decrease.

The third model involves radiation starting from the black hole’s event horizon $r_h$. By examining Fig. \ref{fig.ISSS}, it can be observed that the larger the charge quantity, the smaller the position of the emission intensity peak, and the peak value also slightly decreases. In this model, the observed intensity still primarily comes from direct emission. Similar to the second model, the peaks of the photon sphere and the lens ring overlap, making them difficult to distinguish. However, compared to the first two models, the photon sphere and the lens ring contribute more significantly to the observed intensity. In the appearance of the two-dimensional thin disk, the bright photon sphere and the lens ring overlap to form a luminous area.

In each model, the emission intensity peaks near its start radius ($r_{\mathrm{ISCO}}$, $r_p$, or $r_h$) and rapidly decays. For a distant observer, the direct ring provides the largest contribution to observed intensity, while lensing and photon sphere yield smaller (though non-negligible) contributions. As $Q$ increases, both the peak value and its radial location shift inward.

\section{Compared to the Reissner-Nordström black hole}\label{sec:Comparison}
Setting $l=0$ recovers the Reissner-Nordström solution~\cite{Reissner:1916cle}
\begin{equation}
F_a(r)=1-\frac{2 M}{r}+\frac{Q^2}{r^2}.
\end{equation}
The Lorentz-violating parameter $l$ is tightly constrained and may be extremely close to zero. Here, we compare the charged black hole in the Kalb-Ramond background ($l=0.01$, $l=0.05$ and $l=0.1$) to Reissner-Nordström at $Q=0.3$.

\begin{table}[htbp]
\centering
\caption{The range of $b_c$ for direct, lensing, and photon orbits comparing Reissner-Nordström with Kalb-Ramond ($l=0.01$, $l=0.05$ and $l=0.1$) black holes at $Q=0.3$.}
\label{PRRRR}
\begin{tabular}{|c|c|c|c|}
\hline 
 & Direct $(n<0.75)$ & Lensing $(0.75<n<1.25)$ & Photon $(n>1.25)$\\
\hline
Reissner-Nordström 
& $b_c<4.93142$ or $b_c>6.09956$ 
& $4.93142<b_c<5.10795$ and $5.1499<b_c<6.09956$ 
& $5.10795<b_c<5.1499$
\\ \hline
$l=0.01$ 
& $b_c<4.85905$ or $b_c>5.97711$ 
& $4.85905<b_c<5.0294$ and $5.06918<b_c<5.97711$ 
& $5.0294<b_c<5.06918$
\\ \hline
$l=0.05$ 
& $b_c<4.57161$ or $b_c>5.50608$ 
& $4.57161<b_c<4.71861$ and $4.75055<b_c<5.50608$ 
& $4.71861<b_c<4.75055$
\\ \hline
$l=0.1$ 
& $b_c<4.21697$ or $b_c>4.95635$ 
& $4.21697<b_c<4.33787$ and $4.36176<b_c<4.95635$ 
& $4.33787<b_c<4.36176$
\\
\hline
\end{tabular}
\end{table}

\begin{figure*}[htbp]
	\centering
	\subfigure[Reissner-Nordström: $n$ vs.\ $b_c$.]{
		\begin{minipage}[t]{0.24\linewidth}
			\includegraphics[width=1.8in]{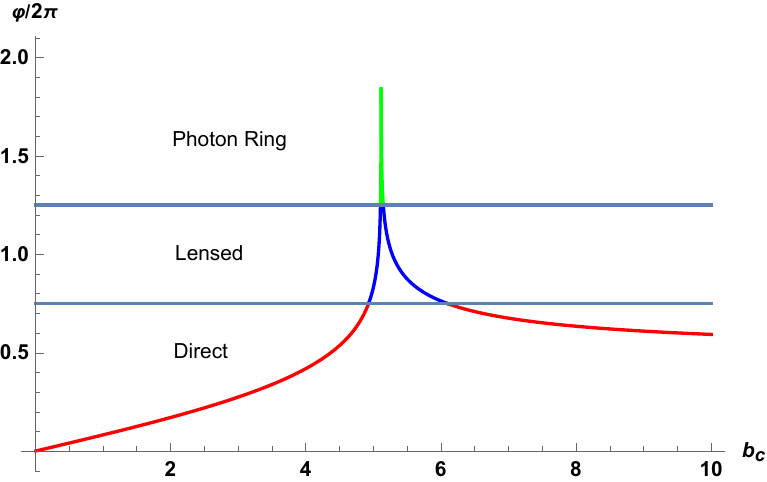}\\
		\end{minipage}
	}
	\subfigure[Reissner-Nordström: photon trajectories.]{
		\begin{minipage}[t]{0.24\linewidth}
			\includegraphics[width=1.8in]{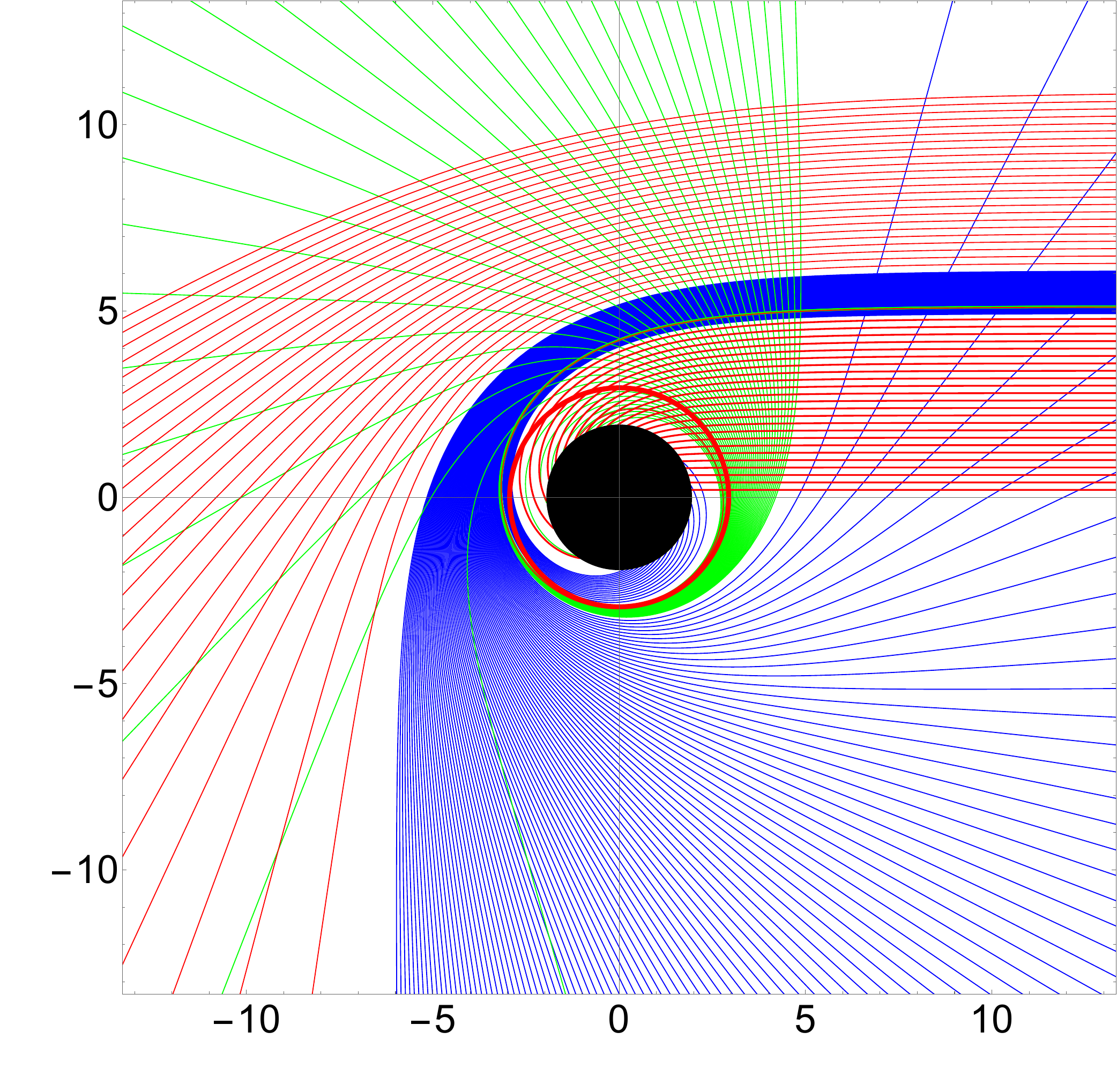}\\
		\end{minipage}
	}
	\subfigure[Reissner-Nordström: transfer function.]{
		\begin{minipage}[t]{0.24\linewidth}
			\includegraphics[width=1.8in]{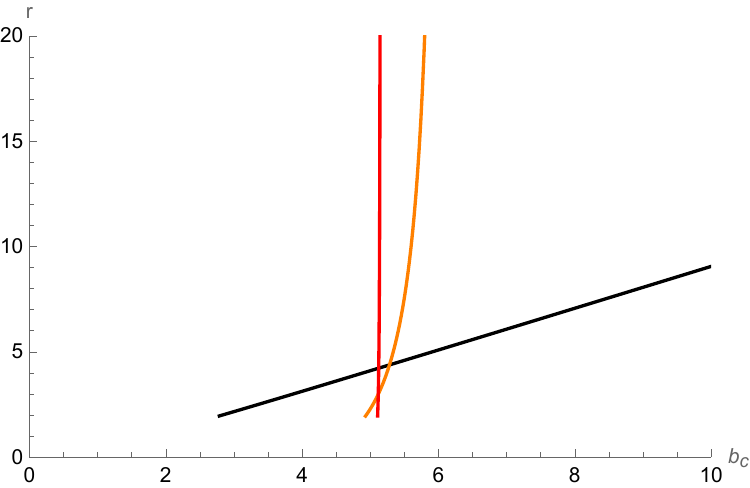}\\
		\end{minipage}
	}
	\\
	\subfigure[Kalb-Ramond($l=0.01$): $n$ vs.\ $b_c$.]{
		\begin{minipage}[t]{0.24\linewidth}
			\includegraphics[width=1.8in]{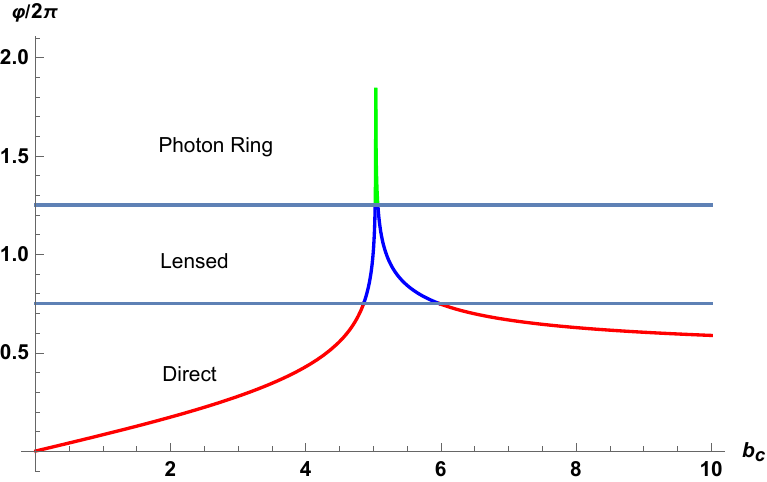}\\
		\end{minipage}
	}
	\subfigure[Kalb-Ramond($l=0.01$): photon trajectories.]{
		\begin{minipage}[t]{0.24\linewidth}
			\includegraphics[width=1.8in]{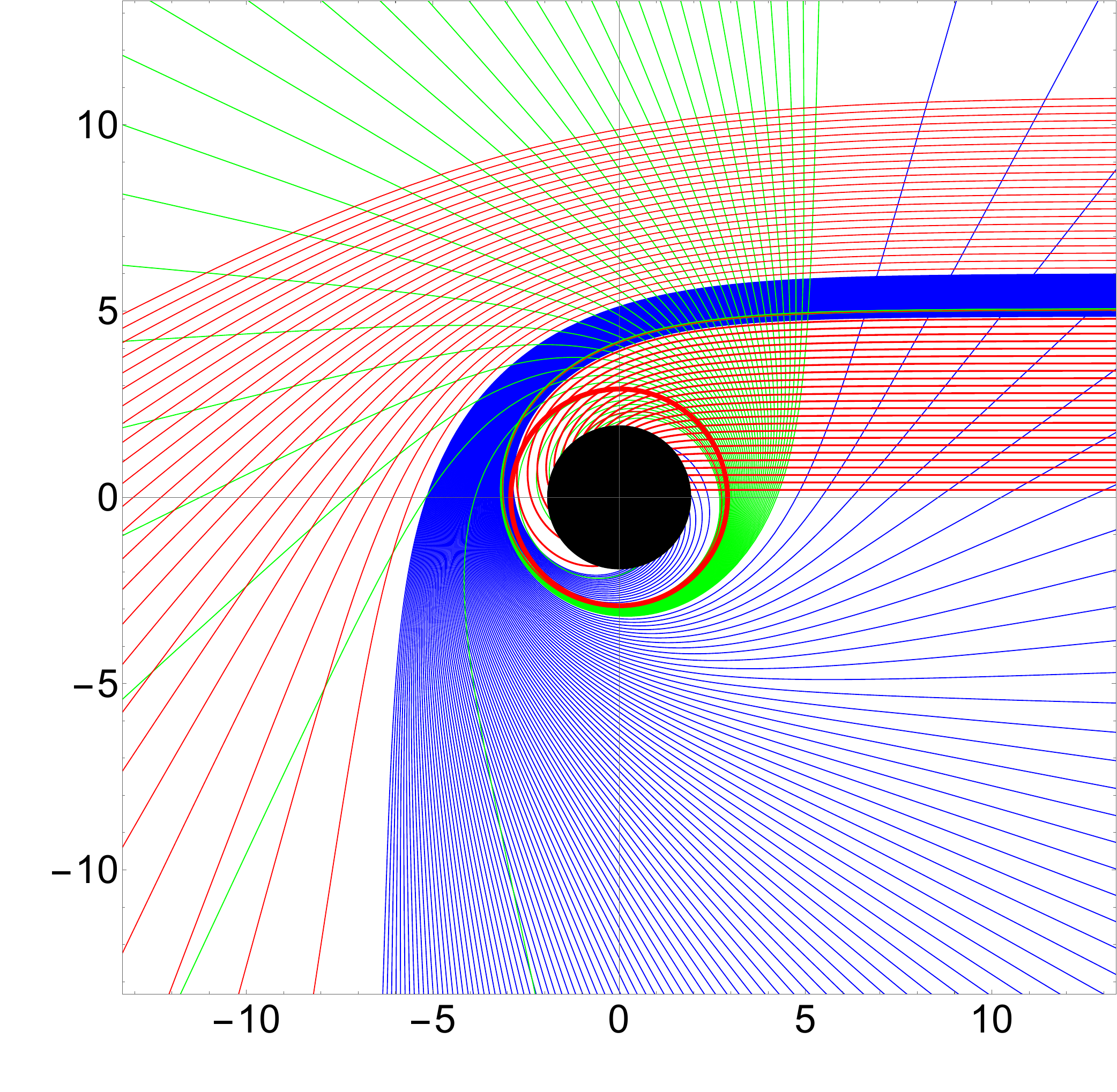}\\
		\end{minipage}
	}
	\subfigure[Kalb-Ramond($l=0.01$): transfer function.]{
		\begin{minipage}[t]{0.24\linewidth}
			\includegraphics[width=1.8in]{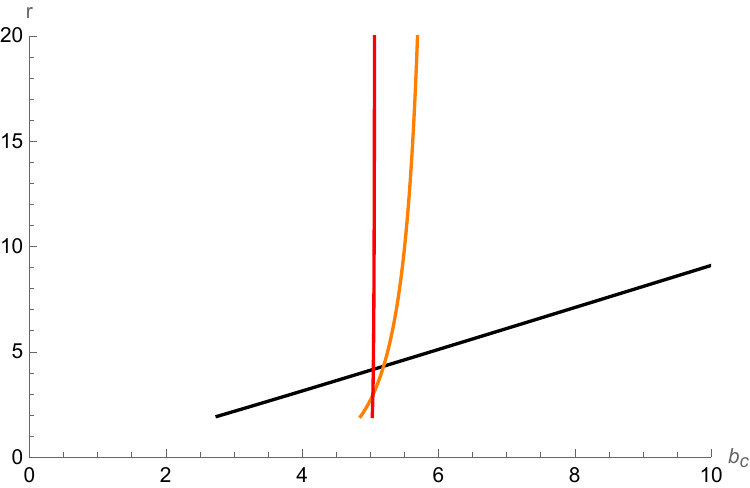}\\
		\end{minipage}
	}
        \\
	\subfigure[Kalb-Ramond($l=0.05$): $n$ vs.\ $b_c$.]{
		\begin{minipage}[t]{0.24\linewidth}
			\includegraphics[width=1.8in]{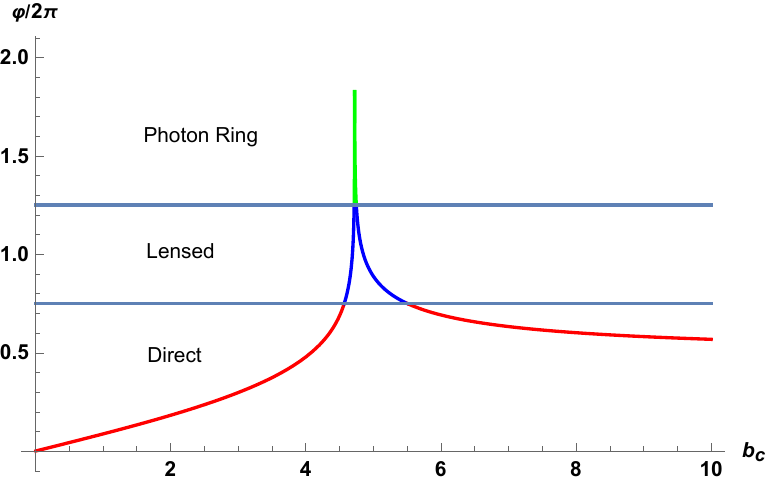}\\
		\end{minipage}
	}
	\subfigure[Kalb-Ramond($l=0.05$): photon trajectories.]{
		\begin{minipage}[t]{0.24\linewidth}
			\includegraphics[width=1.8in]{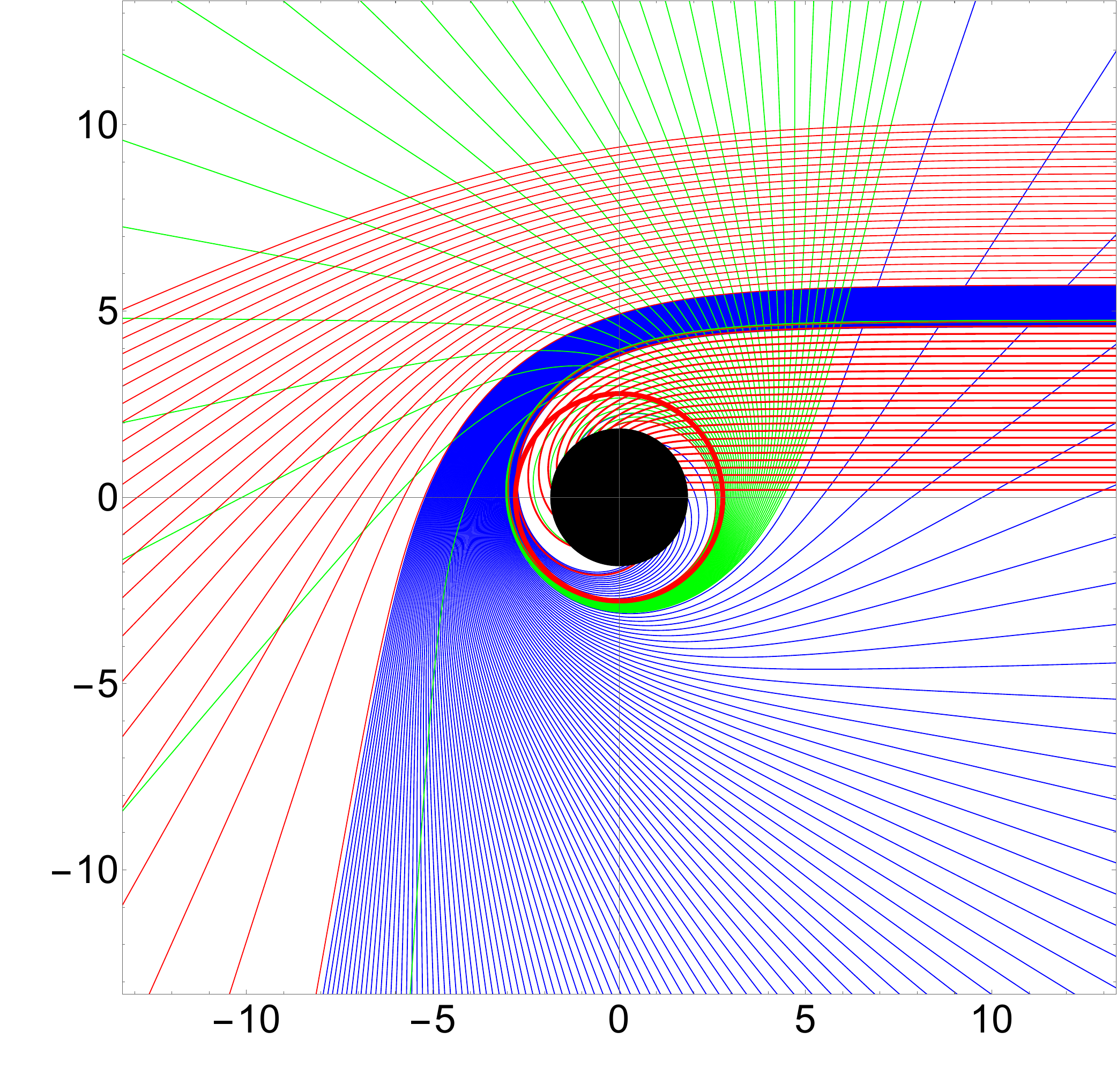}\\
		\end{minipage}
	}
	\subfigure[Kalb-Ramond($l=0.05$): transfer function.]{
		\begin{minipage}[t]{0.24\linewidth}
			\includegraphics[width=1.8in]{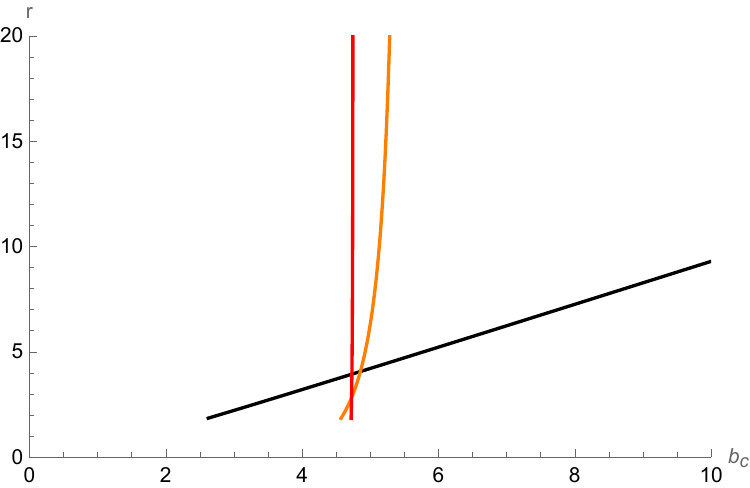}\\
		\end{minipage}
	}
        \\
	\subfigure[Kalb-Ramond($l=0.1$): $n$ vs.\ $b_c$.]{
		\begin{minipage}[t]{0.24\linewidth}
			\includegraphics[width=1.8in]{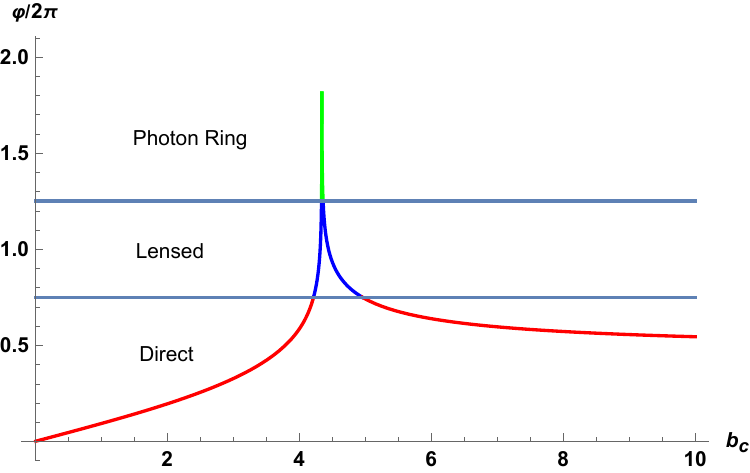}\\
		\end{minipage}
	}
	\subfigure[Kalb-Ramond($l=0.1$): photon trajectories.]{
		\begin{minipage}[t]{0.24\linewidth}
			\includegraphics[width=1.8in]{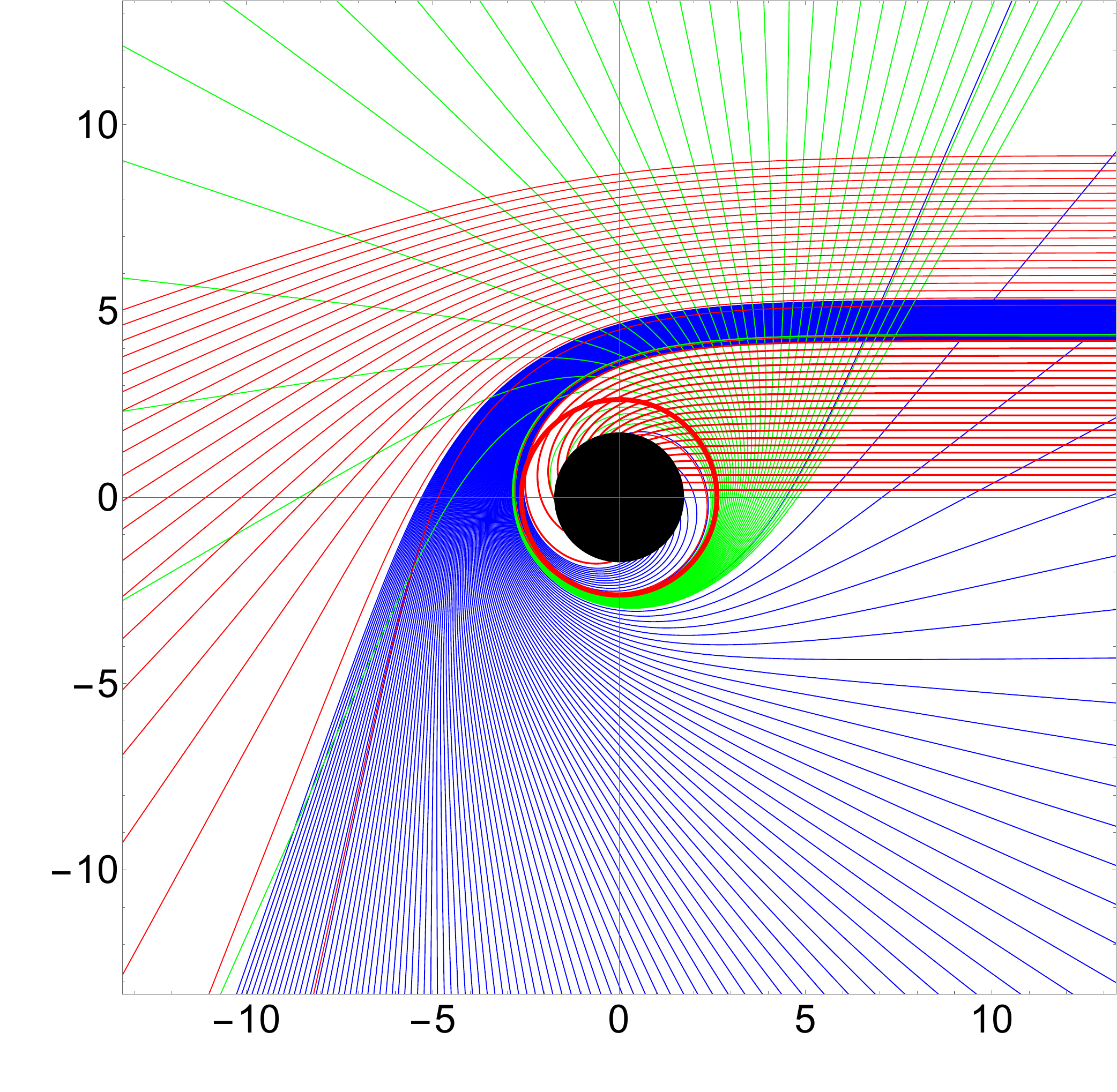}\\
		\end{minipage}
	}
	\subfigure[Kalb-Ramond($l=0.1$): transfer function.]{
		\begin{minipage}[t]{0.24\linewidth}
			\includegraphics[width=1.8in]{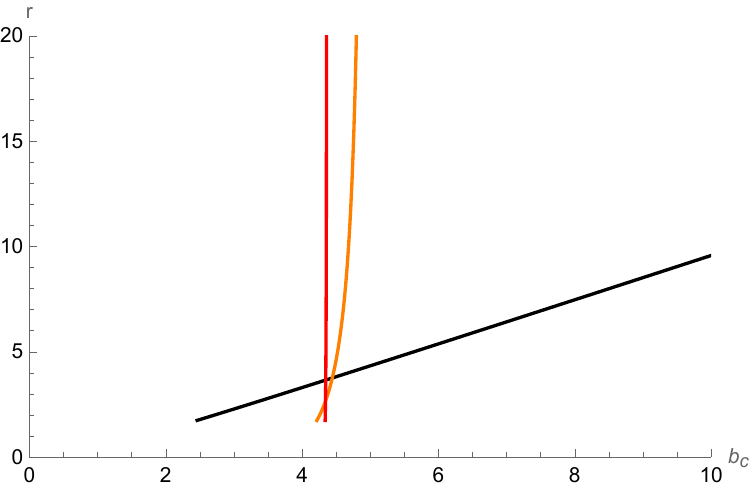}\\
		\end{minipage}
	}
	\caption{(color online) Comparison of Reissner-Nordström ($l=0$) vs.\ Kalb-Ramond ($l=0.01$, $l=0.05$ and $l=0.1$) black holes at $Q=0.3$. The upper row shows $n(b_c)$, photon trajectories, and transfer functions for Reissner-Nordström; The following lines show the same situation for Kalb-Ramond($l=0.01$, $l=0.05$ and $l=0.1$).}
	\label{fig.A}
\end{figure*}

From Table~\ref{tan}, as $l$ increases, $b_h$, $r_p$, and $r_h$ decrease for the Kalb-Ramond black hole. Differently table~\ref{PRRRR} and Fig. \ref{fig.A} demonstrate that the photon sphere and lensing ring for the Kalb-Ramond black hole are narrower.As the Lorentz-violating parameter $l$ increases, the photon and lensing rings of the Kalb-Ramond black hole become progressively narrower, though their thickness slightly decreases. We observe that this behavior is in contrast to the result seen earlier, where an increase in the charge $Q$ led to a different outcome. Lorentz violation enhances the gravitational pull, shrinking these rings. From their transfer functions, direct emission remains the primary contributor. The photon sphere of Reissner-Nordström is more prominent and thus easier to observe.

\begin{figure*}[htbp]
	\centering
	\subfigure[Reissner-Nordström: Models A,B,C.]{
		\begin{minipage}[t]{0.2\linewidth}
			\includegraphics[width=1.5in]{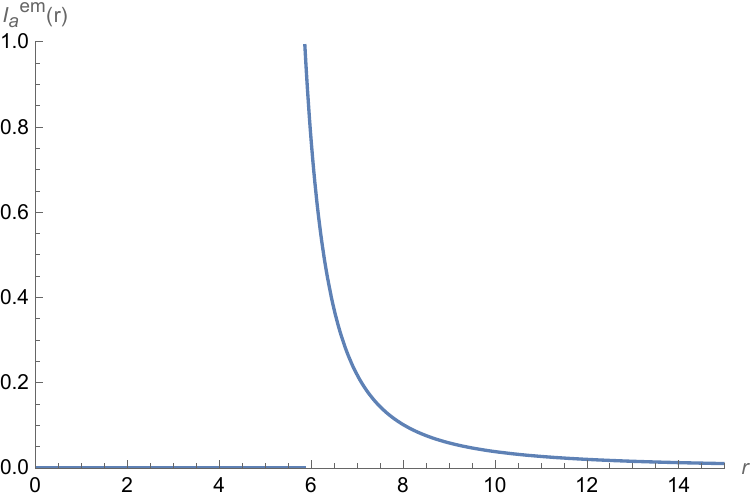}\\
			\includegraphics[width=1.5in]{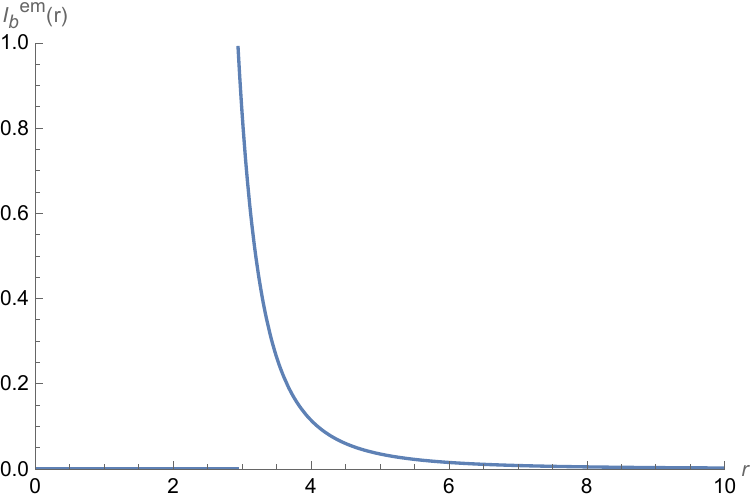}\\
			\includegraphics[width=1.5in]{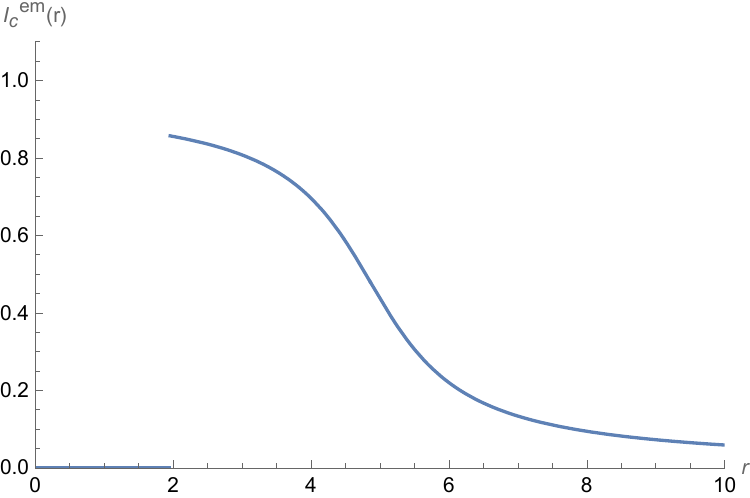}\\
		\end{minipage}
	}
	\subfigure[Reissner-Nordström: Observed intensities.]{
		\begin{minipage}[t]{0.22\linewidth}
			\includegraphics[width=1.5in]{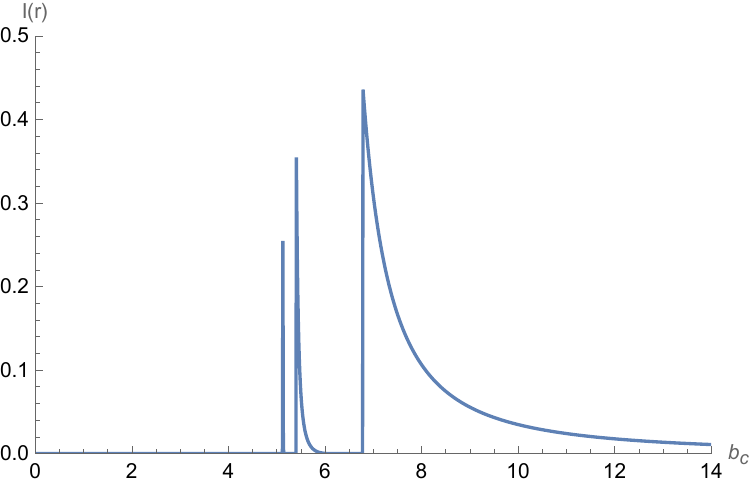}\\
			\includegraphics[width=1.5in]{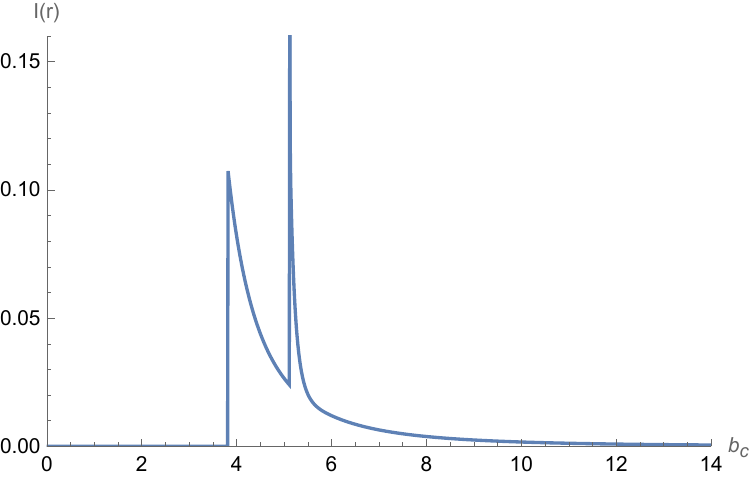}\\
			\includegraphics[width=1.5in]{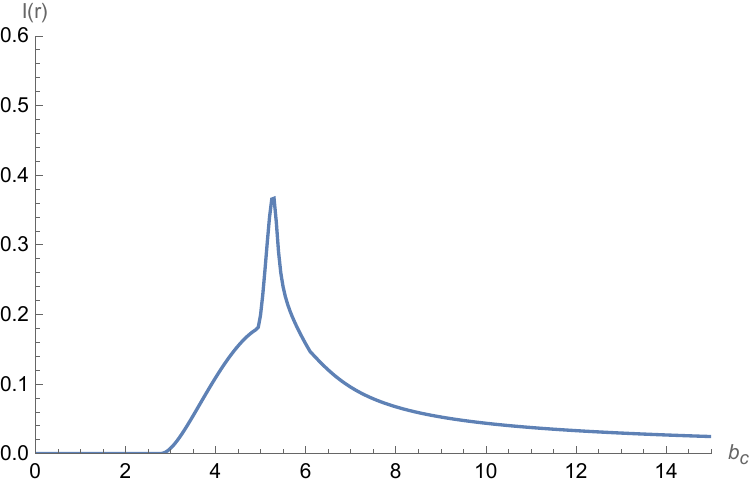}\\
		\end{minipage}
	}
	\subfigure[Kalb-Ramond ($l=0.01$): Models A,B,C.]{
		\begin{minipage}[t]{0.2\linewidth}
			\includegraphics[width=1.5in]{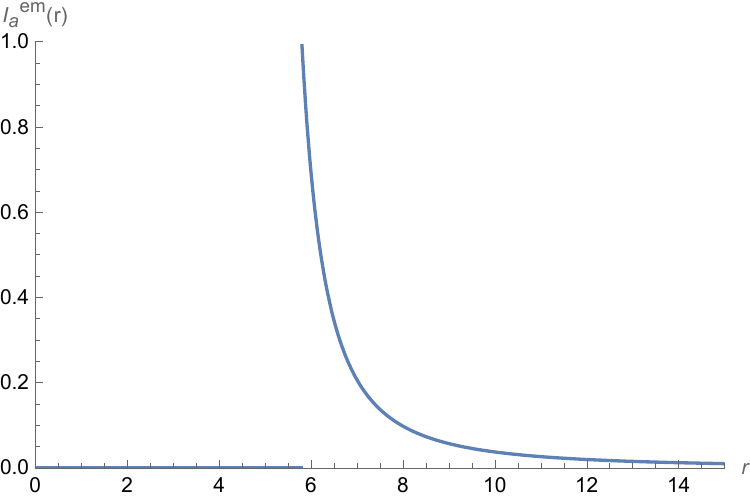}\\
			\includegraphics[width=1.5in]{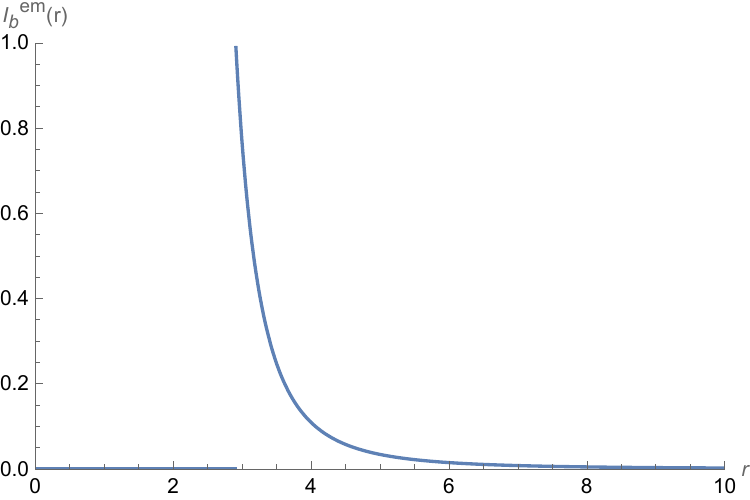}\\
			\includegraphics[width=1.5in]{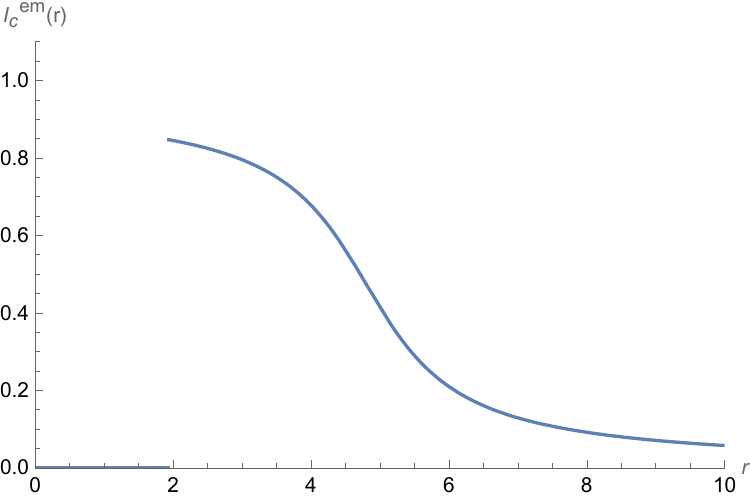}\\
		\end{minipage}
	}  
	\subfigure[Kalb-Ramond ($l=0.01$): Observed intensities.]{
		\begin{minipage}[t]{0.22\linewidth}
			\includegraphics[width=1.5in]{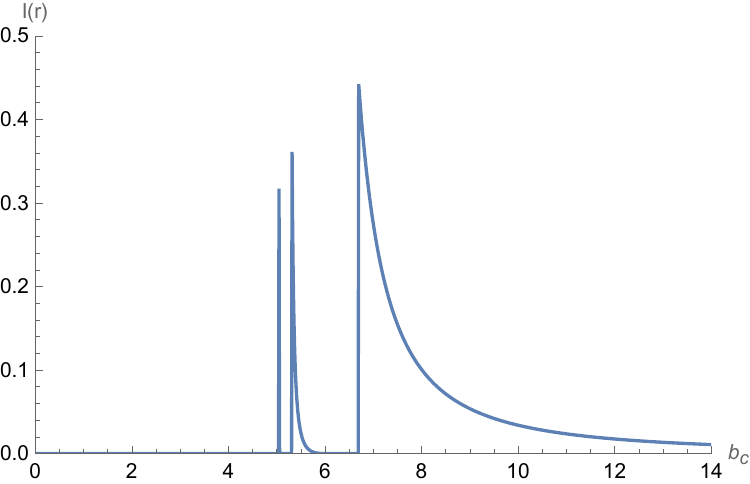}\\
			\includegraphics[width=1.5in]{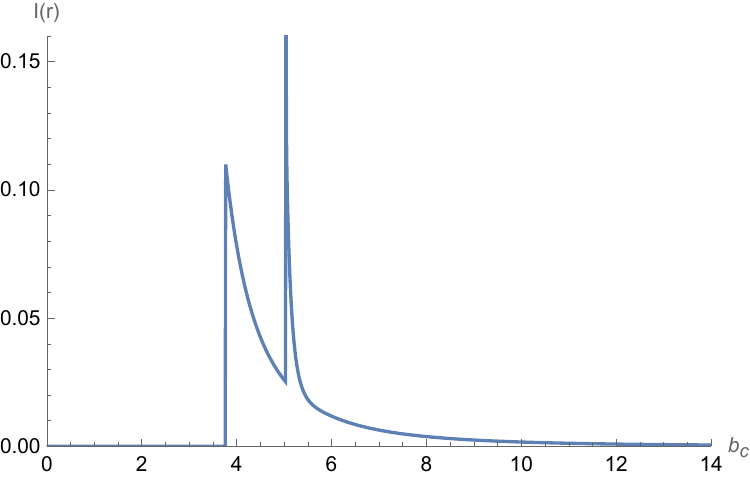}\\
			\includegraphics[width=1.5in]{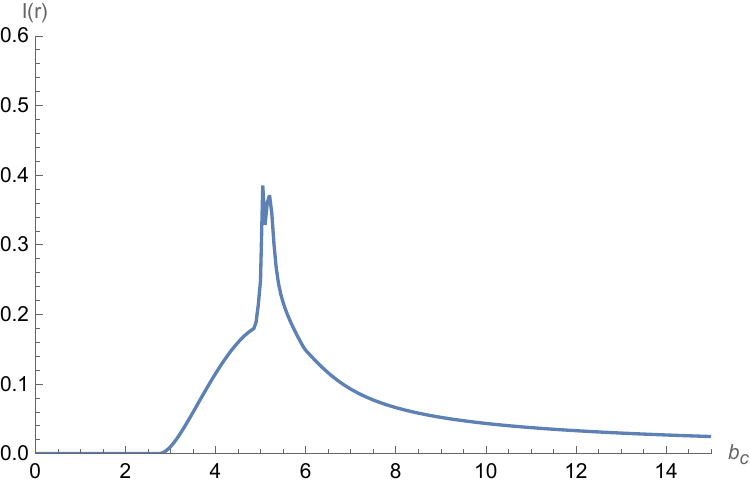}\\
		\end{minipage}
	}
	\caption{(color online) Observational characteristics of a thin accretion disk around Reissner-Nordström ($M=1$, $Q=0.3$) and Kalb-Ramond black hole ($l=0.01$, $M=1$, $Q=0.3$) for Models A, B, and C. The first column and the third column: emission intensity vs.\ $r$. The second column and the fourth column: observed intensity vs.\ $b_c$.}
	\label{fig.AA}
\end{figure*}

\begin{figure*}[htbp]
	\centering
	\subfigure[Kalb-Ramond ($l=0.05$): Models A,B,C.]{
		\begin{minipage}[t]{0.2\linewidth}
			\includegraphics[width=1.5in]{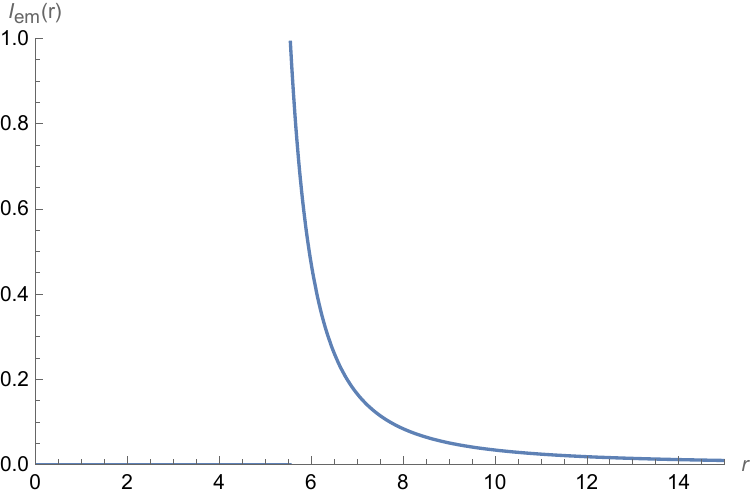}\\
			\includegraphics[width=1.5in]{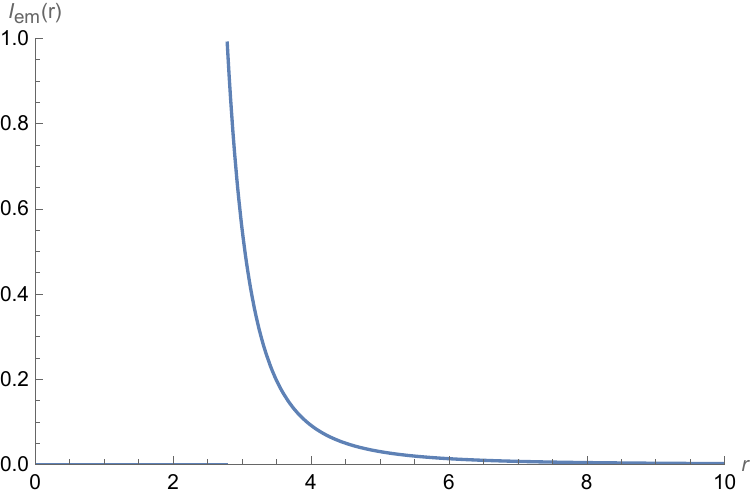}\\
			\includegraphics[width=1.5in]{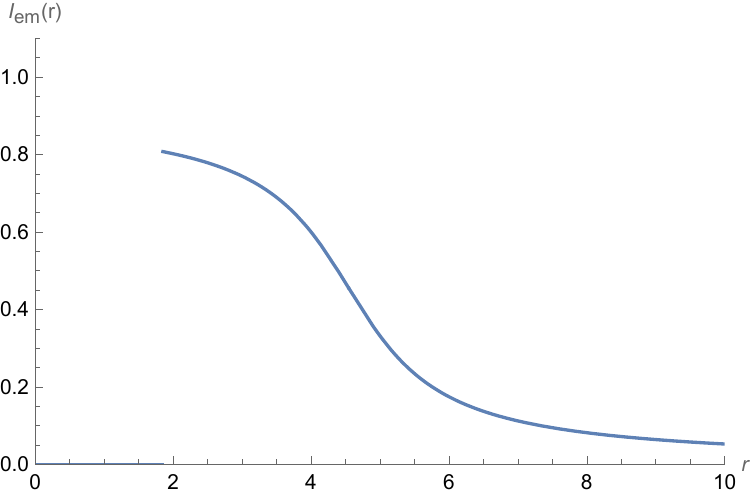}\\
		\end{minipage}
	}
	\subfigure[Kalb-Ramond ($l=0.05$): Observed intensities.]{
		\begin{minipage}[t]{0.2\linewidth}
			\includegraphics[width=1.5in]{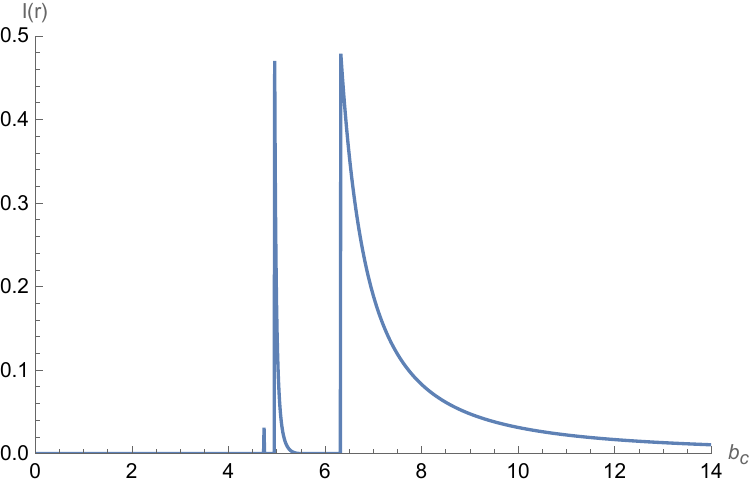}\\
			\includegraphics[width=1.5in]{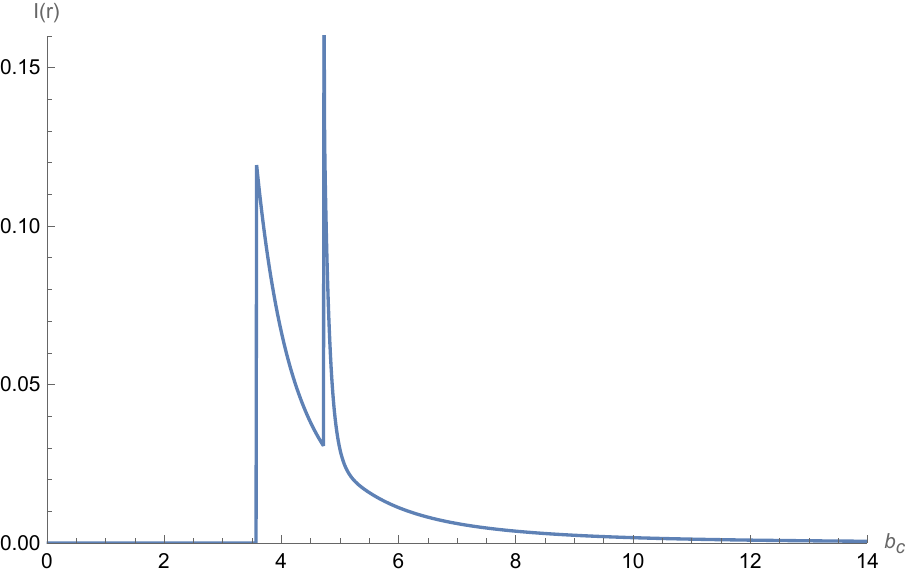}\\
			\includegraphics[width=1.5in]{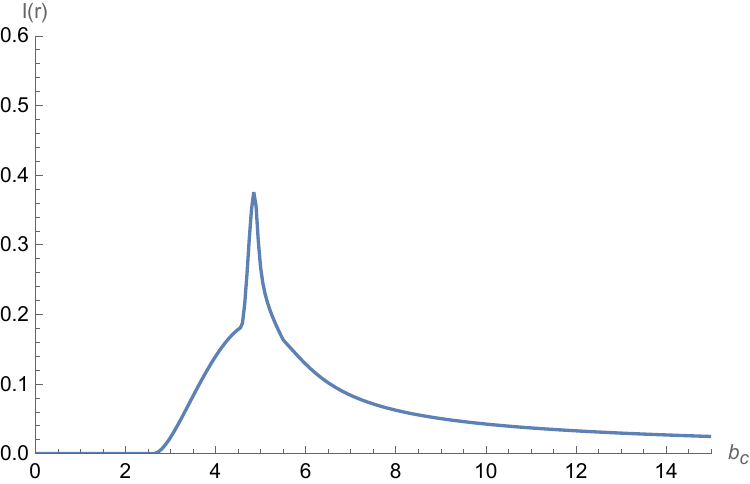}\\
		\end{minipage}
	}
	\subfigure[Kalb-Ramond ($l=0.1$): Models A,B,C.]{
		\begin{minipage}[t]{0.2\linewidth}
			\includegraphics[width=1.5in]{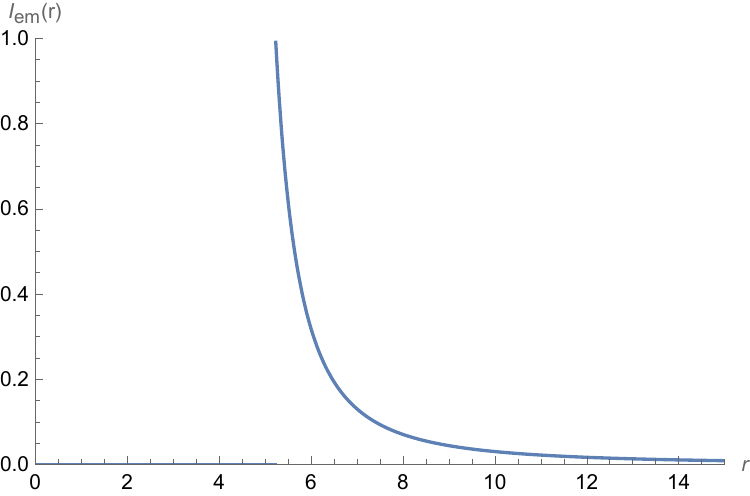}\\
			\includegraphics[width=1.5in]{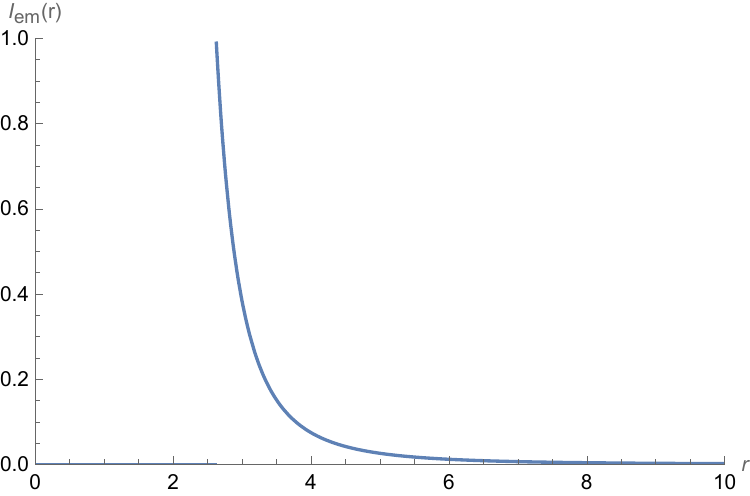}\\
			\includegraphics[width=1.5in]{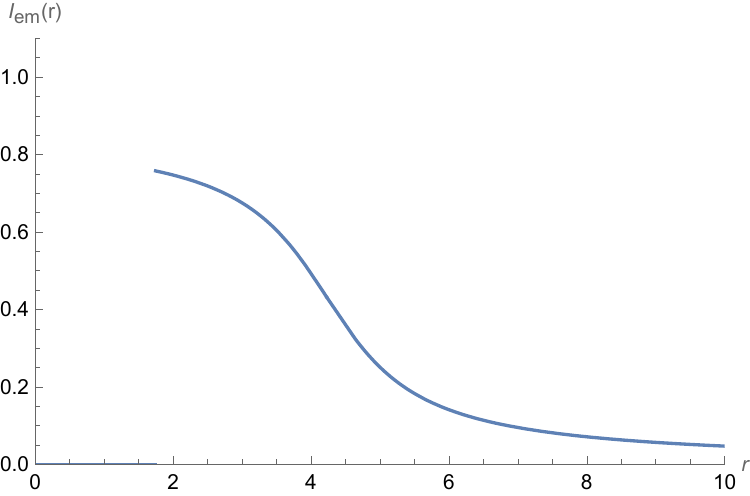}\\
		\end{minipage}
	}
	\subfigure[Kalb-Ramond ($l=0.1$): Observed intensities.]{
		\begin{minipage}[t]{0.2\linewidth}
			\includegraphics[width=1.5in]{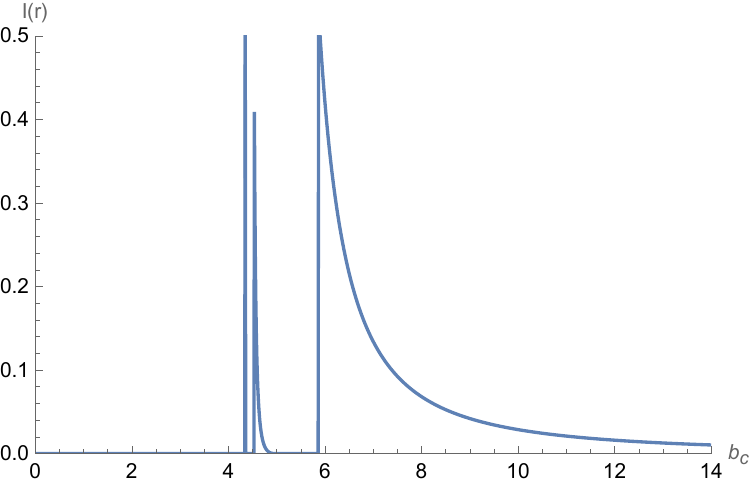}\\
			\includegraphics[width=1.5in]{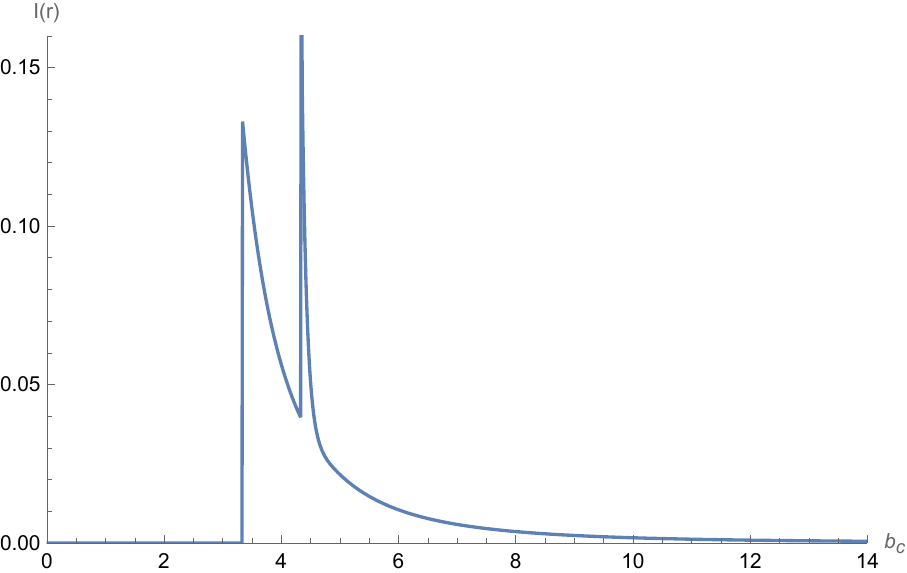}\\
			\includegraphics[width=1.5in]{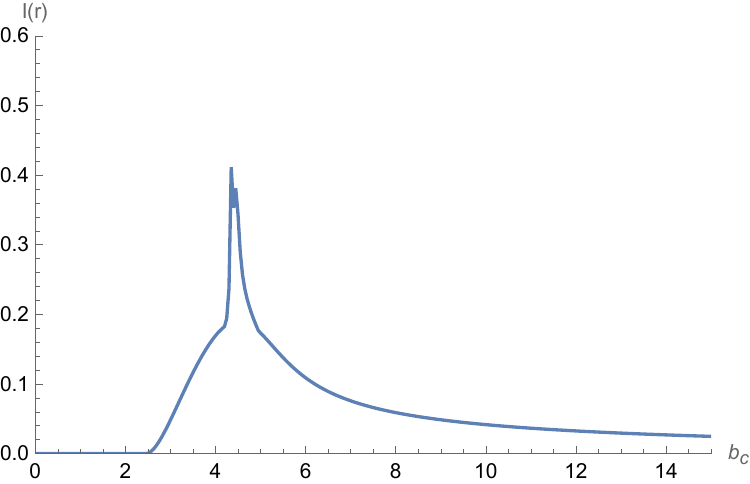}\\
		\end{minipage}
	}
	\caption{(color online) Observational characteristics of a thin accretion disk around a Kalb-Ramond black hole ($l=0.05$, $M=1$, $Q=0.3$) and Kalb-Ramond black hole ($l=0.1$, $M=1$, $Q=0.3$) for Models A, B, and C. Same plotting scheme as Fig.~\ref{fig.AA}.}
	\label{fig.AAA}
\end{figure*}

Figs. \ref{fig.AA} and \ref{fig.AAA} compare the black hole images under three emission models. it is evident that in all three models, the emission intensity reaches a peak and then decreases rapidly. The positions of the peak emission intensity for the Reissner-Nordström black hole in the three models are denoted as $r \approx 5.87 \mathrm{M}$, $r \approx 2.94 \mathrm{M}$, and $r \approx 1.95 \mathrm{M}$, respectively. For the charged black hole in the Kalb-Ramond background with $l=0.01$, the positions of the peak emission intensity in the three models are denoted as $r \approx 5.79 \mathrm{M}$, $r \approx 2.89 \mathrm{M}$, and $r \approx 1.92 \mathrm{M}$, respectively. When $l=0.05$, the corresponding peak positions are located at $r \approx 5.53 \mathrm{M}$, $r \approx 2.78 \mathrm{M}$, and $r \approx 1.82 \mathrm{M}$ respectively. Moreover, when $l=0.1$, the corresponding peak positions respectively change to $r \approx 5.21 \mathrm{M}$, $r \approx 2.62 \mathrm{M}$, and $r \approx 1.73 \mathrm{M}$. Through the above analysis, we can find a pattern, that is, emission intensity of the black hole decreases as Lorentz-violating parameter $l$ increases. In addition, we also find that the peak emission intensity of the charged black hole in the Kalb-Ramond background with $l$ is also slightly less than that of the Reissner-Nordström black hole. In each model, the peak emission intensities for the Lorentz-violating black hole are slightly smaller and occur at smaller radii. This shift indicates that Lorentz-violation leads to reduced emission strength and more compact emission regions.

By comparing the observed intensity images in Figs. \ref{fig.AA} and \ref{fig.AAA}. For the observed intensity distribution of Model $A$, in the case of the Reissner - Nordström black hole, the observed specific intensity reaches its peak at $b_c \approx 5.10 \mathrm{M}$, corresponding to direct emission. A narrower and slightly lower peak appears at $b_c \approx 5.38 \mathrm{M}$, corresponding to the lensing ring. An extremely narrow peak is observed at $b_c \approx 6.75 \mathrm{M}$, which corresponds to the photon sphere. In the charged black hole within the Kalb-Ramond background with $l=0.01$, the corresponding positions shift to $b_c \approx 5.01 \mathrm{M}$, $b_c \approx 5.30 \mathrm{M}$, and $b_c \approx 6.67 \mathrm{M}$, respectively.

When considering the observed intensity distribution of Model $B$, within the Reissner-Nordström black hole, direct emission is found at $\approx 3.78 \mathrm{M}$, whereas the lensing ring makes its appearance at $b_c \approx 5.12 \mathrm{M}$. In the charged black hole within the Kalb-Ramond background with $l=0.01$, these positions shift to $b_c \approx 3.73 \mathrm{M}$ and $b_c \approx 5.02 \mathrm{M}$, respectively. The emission regions of the photon sphere and the lensing ring exhibit an overlapping pattern, and the peak intensity of the lensing ring is greater than that of the direct emission.

In Model $C$, within the Reissner-Nordström black hole, the observed intensity distribution reveals that direct emission emerges around $b_c \approx 2.85 \mathrm{M}$, while the lensing ring appears at approximately $b_c \approx 5.26 \mathrm{M}$, overlapping with the direct emission. However, in a charged black hole influenced by the Kalb-Ramond background with $l=0.01$, these positions shift to $b_c \approx 2.79 \mathrm{M}$ and $b_c \approx 5.07 \mathrm{M}$, respectively. As the Lorentz-violation parameter $l$ increases, both the peak intensity and its corresponding position gradually decrease.

Hence, Lorentz violation leads to smaller photon sphere radii, event horizon radii, and ISCO radii, all shifting inward. Under the same emission model, both the peak emiission intensity and its location move to smaller values, and the impact parameter ranges for the photon sphere, lensing ring, and direct ring diminish, suggesting that black holes with Lorentz violation are more difficult to detect.

\section{Conclusion and Discussion}\label{sec:Conclusion}
Because black holes cannot be observed directly, surrounding accretion material is crucial. This paper analyzed the optical appearance of a charged black hole in the Kalb-Ramond background, focusing on Lorentz violation with $l=0.01$.

In a spherically symmetric setup, we examined the null geodesics, derived the effective potential, and computed the photon sphere radii, event horizon radii, and critical impact parameter. Under $l=0.01$, these quantities decrease with increasing $Q$. Classifying photons by the number of equatorial-plane intersections yielded three trajectory types: direct, lensing, and photon orbits. Ray-tracing allowed us to determine the corresponding $b_c$ ranges, and the transfer function analysis indicated that direct emission dominates the total observed intensity, with lensing ring and photon sphere contributing significantly less.

Finally, we explored black hole shadows under three emission models, each time finding that the emission intensity peaks at the starting radius and then falls off sharply. For an observer at infinity, direct emission is again the principal source of brightness. As the values of $l$ and $Q$ increase, both the emission peak and the observed intensity peak shift inward, and their positions slightly decrease. Comparing to the Reissner-Nordström black hole, we confirmed that the Lorentz-violation parameter brings $r_P$, $r_h$, and $r_{\mathrm{ISCO}}$ closer to the center, reducing and shifting the intensity peaks. These results suggest that black holes with Lorentz violation are comparatively more difficult to detect observationally due to their diminished and more compact emission characteristics.

In this paper, we employ three radiation models ($A$, $B$, and $C$) to investigate the radiative properties of optically thin accretion disks. However, for optically thick accretion disks, the significant optical depth leads to highly complex photon trajectories within the disk, involving multiple scattering and absorption events. This substantially increases the computational cost of numerical simulations, posing significant challenges for accurately modeling the radiation field. In future research, more advanced radiation models, such as general relativistic radiative transfer and Monte Carlo simulations, can be utilized to further explore the effects of multiple scattering and optical depth on black hole shadows and photon rings. These refined methodologies will provide deeper insights, enabling a better understanding of the observational characteristics of black holes and the radiative properties of accretion disks.

To determine the critical value of the parameter $l$, we can discuss it by using the shadow of the $M87$ black hole detected by the Event Horizon Telescope. The diameter of the $M87$ black-hole shadow in units of mass has been obtained, that is $d_{M 87} \equiv D \cdot \delta / M \approx 11 \pm 1.5$,  where $D$ and $\delta$ the angular size and the distance of the $M87$ black-hole shadow respectively~\cite{Allahyari:2019jqz,Jafarzade:2021umv,Bambi:2019tjh,Li:2021ypw}. Within the uncertainties of $1\delta$ and $2\delta$, it is easy to determine that the regions where the shadow diameter lies are $9.5 \sim 12.5$ and $8 \sim 14$. Therefore, this range may give some upper or lower limits to the black-hole parameters.
We assume that when $Q=0.3$, under the condition of an uncertainty of $1\delta$, we can calculate that the range of the parameter $l$ is between $-0.1423$ and $0.04687$. While under the condition of an uncertainty of $2\delta$, the corresponding value of the parameter $l$ lies between $-0.2449$ and $0.1486$.In summary, the shadow of a black hole can effectively place certain observational constraints on its relevant parameters. This is of profound significance for delving deeper into the appearance characteristics of black holes in future observational studies.

This paper primarily investigates the optical properties of charged black holes in the presence of a Kalb-Ramond field.  Kalb-Ramond, originating from string theory, is a rank-two antisymmetric tensor field and constitutes an essential component of the theory. However, string theory also introduces other modifications, such as Dilaton fields. The introduction of the Dilaton field further modifies the geometry and thermodynamic quantities of black holes, playing a crucial role in string theory. Although the roles of the Dilaton field and the Kalb-Ramond field differ in string theory, the methodologies employed in this study to analyze the Kalb-Ramond field—such as null geodesic analysis and ray-tracing techniques-can be extended to models incorporating the Dilaton field. The dynamical effects of the Dilaton field influence the geometry and topology of black holes, potentially affecting observable quantities such as the photon sphere, black hole shadow, and accretion disk. Therefore, the introduction of the Dilaton field may alter the shape and size of the black hole shadow, similar to the Lorentz-violating effects induced by the Kalb-Ramond field. Future research can further explore the impact of the Dilaton field on the photon sphere and black hole shadow, comparing the results with our findings to gain deeper insights into these effects.

\section*{Acknowledgments}
 This work is supported by the Sichuan Science and Technology Program (2024NSFSC1999).



\begin{thebibliography}{99}

\bibitem{Schwarzschild:1916uq}
K.~Schwarzschild,
Sitzungsber. Preuss. Akad. Wiss. Berlin (Math. Phys. ) \textbf{1916}, 189 (1916).

\bibitem{Kerr:1963ud}
R.~P.~Kerr,
Phys. Rev. Lett. \textbf{11},237 (1963).

\bibitem{Narayan:2019imo}
R.~Narayan, M.~D.~Johnson and C.~F.~Gammie,
Astrophys. J. Lett. \textbf{885}, L33 (2019).

\bibitem{Zakharov:2023yjl}
A.~F.~Zakharov,
Int. J. Mod. Phys. D \textbf{33}, 2340004 (2024).

\bibitem{Safarzadeh:2019imq}
M.~Safarzadeh, A.~Loeb and M.~Reid,
Mon. Not. Roy. Astron. Soc. \textbf{488}, L90 (2019).

\bibitem{Zeng:2020vsj}
X.~X.~Zeng and H.~Q.~Zhang,
Eur. Phys. J. C \textbf{80}, 1058 (2020).

\bibitem{He:2024bll}
K.~J.~He, Y.~W.~Han and G.~P.~Li,
Phys. Dark Univ. \textbf{44}, 101468 (2024).

\bibitem{Wei:2021vdx}
S.~W.~Wei and Y.~X.~Liu,
Phys. Rev. D \textbf{105}, 104003 (2022).

\bibitem{Synge:1966okc}
J.~L.~Synge,
Mon. Not. Roy. Astron. Soc. \textbf{131}, 463 (1966).

\bibitem{Bardeen:1973tla}
J.~M.~Bardeen,
Proceedings, Ecole d'Et\'e de Physique Th\'eorique: Les Astres Occlus : Les Houches, France, August, 1972, 215 (1973).

\bibitem{He:2022yse}
K.~J.~He, S.~C.~Tan and G.~P.~Li,
Eur. Phys. J. C \textbf{82}, 81 (2022).

\bibitem{He:2021htq}
K.~J.~He, S.~Guo, S.~C.~Tan and G.~P.~Li,
Chin. Phys. C \textbf{46}, 085106 (2022).

\bibitem{Jin:2020emq}
X.~H.~Jin, Y.~X.~Gao and D.~J.~Liu,
Int. J. Mod. Phys. D \textbf{29}, 2050065 (2020).

\bibitem{Perlick:2021aok}
V.~Perlick and O.~Y.~Tsupko,
Phys. Rept. \textbf{947}, 1 (2022).

\bibitem{Mustafa:2022xod}
G.~Mustafa, F.~Atamurotov, I.~Hussain, S.~Shaymatov and A.~\"Ovg\"un,
Chin. Phys. C \textbf{46}, 125107 (2022).

\bibitem{Guo:2020zmf}
M.~Guo and P.~C.~Li,
Eur. Phys. J. C \textbf{80}, 588 (2020).

\bibitem{Chen:2019fsq}
Y.~Chen, J.~Shu, X.~Xue, Q.~Yuan and Y.~Zhao,
Phys. Rev. Lett. \textbf{124}, 061102 (2020).

\bibitem{Zeng:2021mok}
X.~X.~Zeng, K.~J.~He and G.~P.~Li,
Sci. China Phys. Mech. Astron. \textbf{65}, 290411 (2022).

\bibitem{He:2024amh}
K.~J.~He, G.~P.~Li, C.~Y.~Yang and X.~X.~Zeng, arXiv:2411.11680 [astro-ph.HE] (2024).

\bibitem{Yang:2024nin}
C.~Y.~Yang, M.~I.~Aslam, X.~X.~Zeng and R.~Saleem, arXiv:2411.11807 [astro-ph.HE] (2024).

\bibitem{Gralla:2019xty}
S.~E.~Gralla, D.~E.~Holz and R.~M.~Wald,
Phys. Rev. D \textbf{100}, 024018 (2019).

\bibitem{Li:2021ypw}
G.~P.~Li and K.~J.~He,
Eur. Phys. J. C \textbf{81}, 1018 (2021).

\bibitem{Bronzwaer:2021lzo}
T.~Bronzwaer and H.~Falcke,
Astrophys. J. \textbf{920}, 155 (2021).

\bibitem{Gan:2021xdl}
Q.~Gan, P.~Wang, H.~Wu and H.~Yang,
Phys. Rev. D \textbf{104}, 044049 (2021).

\bibitem{EventHorizonTelescope:2021dqv}
P.~Kocherlakota \textit{et al.} [Event Horizon Telescope],
Phys. Rev. D \textbf{103}, 104047 (2021).

\bibitem{Gan:2021pwu}
Q.~Gan, P.~Wang, H.~Wu and H.~Yang,
Phys. Rev. D \textbf{104}, 024003 (2021).

\bibitem{Wang:2023vcv}
X.~J.~Wang, X.~M.~Kuang, Y.~Meng, B.~Wang and J.~P.~Wu,
Phys. Rev. D \textbf{107}, 124052 (2023).

\bibitem{Uniyal:2023inx}
A.~Uniyal, S.~Chakrabarti, R.~C.~Pantig and A.~\"Ovg\"un,
New Astron. \textbf{111}, 102249 (2024).

\bibitem{Junior:2024ety}
E.~L.~B.~Junior, J.~T.~S.~S.~Junior, F.~S.~N.~Lobo, M.~E.~Rodrigues, D.~Rubiera-Garcia, L.~F.~D.~da Silva and H.~A.~Vieira,
Eur. Phys. J. C \textbf{84}, 1257 (2024).

\bibitem{Zhang:2025wnu}
X.~X.~Zhang and Y.~H.~Jiang,
Chin. Phys. C \textbf{49}, 025101(2025).

\bibitem{Nandi:2023vxt}
K.~K.~Nandi, R.~N.~Izmailov, R.~K.~Karimov and A.~A.~Potapov,
Eur. Phys. J. C \textbf{83}, 984 (2023).

\bibitem{Jusufi:2017hed}
K.~Jusufi, I.~Sakall\i{} and A.~\"Ovg\"un,
Phys. Rev. D \textbf{96}, 024040 (2017).

\bibitem{Kanzi:2019gtu}
S.~Kanzi and \.I.~Sakall\i{},
Nucl. Phys. B \textbf{946}, 114703 (2019).

\bibitem{Ditta:2024lnb}
A.~Ditta, F.~Javed, A.~Bouzenada, G.~Mustafa, A.~Mahmood, F.~Atamurotov and V.~Khamidov,
JHEAp \textbf{45}, 62 (2025).

\bibitem{Ma:2024ets}
Y.~Ma, S.~Zheng, H.~Li and B.~Li,
Nucl. Phys. B \textbf{1009}, 116732 (2024).

\bibitem{Zahid:2024hyy}
M.~Zahid, F.~Sarikulov, C.~Shen, J.~Rayimbaev, K.~Badalov and S.~Muminov,
Chin. J. Phys. \textbf{91}, 45 (2024).

\bibitem{al-Badawi:2024pdx}
A.~al-Badawi, S.~Shaymatov and I.~Sakall\i{},
Eur. Phys. J. C \textbf{84}, 825 (2024).

\bibitem{Jumaniyozov:2025wcs}
S.~Jumaniyozov, M.~Zahid, M.~Alloqulov, I.~Ibragimov, J.~Rayimbaev and S.~Murodov,
Eur. Phys. J. C \textbf{85}, 126 (2025).

\bibitem{Liu:2024oas}
W.~Liu, D.~Wu and J.~Wang,
JCAP \textbf{09}, 017 (2024).

\bibitem{Ortiqboev:2024mtk}
D.~Ortiqboev, F.~Javed, F.~Atamurotov, A.~Abdujabbarov and G.~Mustafa,
Phys. Dark Univ. \textbf{46}, 101615 (2024).

\bibitem{Raza:2024zkp}
M.~A.~Raza, M.~Zubair and E.~Maqsood,
JCAP \textbf{05}, 047 (2024).

\bibitem{Junior:2024vdk}
E.~L.~B.~Junior, J.~T.~S.~S.~Junior, F.~S.~N.~Lobo, M.~E.~Rodrigues, D.~Rubiera-Garcia, L.~F.~D.~da Silva and H.~A.~Vieira,
Phys. Rev. D \textbf{110}, 024077 (2024).

\bibitem{Atamurotov:2022wsr}
F.~Atamurotov, D.~Ortiqboev, A.~Abdujabbarov and G.~Mustafa,
Eur. Phys. J. C \textbf{82}, 659 (2022).

\bibitem{Kumar:2020hgm}
R.~Kumar, S.~G.~Ghosh and A.~Wang,
Phys. Rev. D \textbf{101}, 104001 (2020).

\bibitem{Mavromatos:2018drr}
N.~E.~Mavromatos and S.~Sarkar,
Phys. Rev. D \textbf{97}, 125010 (2018).

\bibitem{Lessa:2019bgi}
L.~A.~Lessa, J.~E.~G.~Silva, R.~V.~Maluf and C.~A.~S.~Almeida,
Eur. Phys. J. C \textbf{80}, 335 (2020).

\bibitem{Altschul:2009ae}
B.~Altschul, Q.~G.~Bailey and V.~A.~Kostelecky,
Phys. Rev. D \textbf{81}, 065028 (2010).

\bibitem{Duan:2023gng}
Z.~Q.~Duan, J.~Y.~Zhao and K.~Yang,
Eur. Phys. J. C \textbf{84}, 798 (2024).

\bibitem{Reissner:1916cle}
H.~Reissner,
Annalen Phys. \textbf{355}, 106 (1916).

\bibitem{Bambi:2019tjh}
C.~Bambi, K.~Freese, S.~Vagnozzi and L.~Visinelli,
Phys. Rev. D \textbf{100}, 044057 (2019).

\bibitem{Jafarzade:2021umv}
K.~Jafarzade, M.~Kord Zangeneh and F.~S.~N.~Lobo,
Annals Phys. \textbf{446}, 169126 (2022).

\bibitem{Allahyari:2019jqz}
A.~Allahyari, M.~Khodadi, S.~Vagnozzi and D.~F.~Mota,
JCAP \textbf{02},  003 (2020).


\end{thebibliography}
\end{document}